\documentclass[reprint,superscriptaddress,amsmath,amssymb,aps,prb,floatfix]{revtex4-2}
\usepackage{xcolor}
\usepackage{graphicx}% Include figure files
\usepackage{dcolumn}% Align table columns on decimal point
\usepackage{bm}% bold math
\usepackage{bbold}
\usepackage{placeins}
\usepackage{nicefrac}
\usepackage{MnSymbol}
\usepackage[breaklinks=true,colorlinks,citecolor=blue,linkcolor=blue,urlcolor=blue]{hyperref}% add hypertext capabilities
\usepackage[percent]{overpic}
\usepackage{changes}
\usepackage{tikz}
\setaddedmarkup{\textcolor{blue}{#1}}                     % Blue for \added text
\setdeletedmarkup{\textcolor{red}{\sout{#1}}}             % Red for \deleted text (optional, with strike-through)

\definechangesauthor[name=Giovanni Citeroni, color=purple]{GC}

\def\a{\alpha}          \def\g{\gamma}   \def\d{\delta} 
          
       \def\l{\lambda}       
                    
\def\t{\tau}           % \def\c{\chi}
        \def\o{\omega}   
\def\G{\Gamma}       \def\D{\Delta}    \def\L{\Lambda}  
                     
        \def\O{\Omega}   

\def\MM{{\cal M}} 
\def\HH{{\cal H}}
\def\NN{{\cal N}}

\def\=={\equiv}

\def\qed{\raise1pt\hbox{\vrule height5pt width5pt depth0pt}}

\def\cG0{{\cal G}_0} 
\def\cG{{\cal G}}

 \def\=={\equiv}

 \def\ep0{\epsilon_{p}} \def\ed0{\epsilon_{f}}

\def\be{\begin{equation}}
\def\ee{\end{equation}}

\protected\def\annh{\phantom{\dagger}}

\def\ac{\hat{a}^{\dagger}}
\def\aa{\hat{a}^{\annh}}
\def\bc{\hat{b}^{\dagger}}
\def\ba{\hat{b}^{\annh}}
\def\gc{\hat{\gamma}^{\dagger}}
\def\ga{\hat{\gamma}^{\annh}}

\newcommand{\ket}[1]{|{#1}\rangle}

\newcommand{\braket}[3]{\langle{#1}| {#2} |{#3} \rangle}

\newcommand{\quave}[1]{\langle{#1}\rangle}

\def\tent{\tau_{\rm ent}}
\def\tw{\tau_{\rm w}}
\def\tsep{\tau_{\rm sep}}
\def\went{w_{\rm ent}}
\def\gonepht{g_{\rm pht}^{(1)}}
\def\gonemat{g_{\rm mat}^{(1)}}

\def\yi{y_\mathrm{i}}
\def\yf{y_\mathrm{f}}
\def\ti{t_\mathrm{i}}
\def\tf{t_\mathrm{f}}
\def\de{\mathrm{d}}
\def\im{\mathrm{i}}
\def\KK{{\cal K}}

\begin{document}

\title{Ultrafast dynamics of quantum matter driven by time-energy entangled photons}
\author{Giovanni Citeroni}
\affiliation{Dipartimento di Fisica dell'Universit\`a di Pisa, Largo Bruno Pontecorvo 3, I-56127 Pisa,~Italy}
\author{Marco Polini}
\affiliation{Dipartimento di Fisica dell'Universit\`a di Pisa, Largo Bruno Pontecorvo 3, I-56127 Pisa,~Italy}
\author{Michael Dapolito}
\affiliation{Department of Physics, Columbia University, New York, NY, 10027, USA}
\author{D. N. Basov}
\affiliation{Department of Physics, Columbia University, New York, NY, 10027, USA}
\author{Giacomo Mazza}
\affiliation{Dipartimento di Fisica dell'Universit\`a di Pisa, Largo Bruno Pontecorvo 3, I-56127 Pisa,~Italy}
\begin{abstract}

We study the dynamics of quantum matter interacting with time-energy entangled photons.
We consider the stimulation of a collective mode of a two-dimensional material by means 
of one of the two partners of a time-energy entangled pair of photons.
Using an exactly solvable model, we analyze the out-of-equilibrium properties of 
both light {\it and} matter degrees of freedom, and show how  entanglement in the 
incident photons deeply modifies relevant time scales of the light-matter interaction 
process.
We find that entanglement strongly suppresses the delay between the transmission and 
absorption events, which become synchronous in the limit of strongly entangled wave packets. 
By comparing numerical simulations with analytic modeling, we trace back 
this behavior to the representation of entangled wave packets in terms 
of a superposition of multiple train pulses containing an increasing 
number of ultrashort non-entangled packets.
As a result, we show that the entangled driving allows the creation 
of a matter excitation on a time scale shorter than the temporal width 
of the pulse. Eventually, by analyzing temporal correlations of the 
excited matter degrees of freedom, we show that driving with entangled photons imprints  
characteristic temporal correlations of time-energy entangled modes 
in the matter degree of freedom.
\end{abstract}
\maketitle
\section{Introduction}
Light stimulation plays a pivotal role in the investigation of materials 
properties. It offers unique information on the 
dynamics of the microscopic degrees of freedom, and, 
at the same time, it promises tantalizing perspectives 
for the creation of transient states of matter not 
achievable in equilibrium 
conditions~\cite{afanasiev_magnetic_interaction_NatMat2021,
fava_field_expulsion2024,mitrano_k3c60,nova_STO_science2019,Li_light_induced_STO_Science2019}.
Traditionally, light stimulation involves the use 
of classical pulses with a macroscopically large number
of photons. In recent years, the investigation of 
light-matter interactions involving purely quantum 
aspects of light have gained a great deal of attention.
For example, in the context of quantum materials 
embedded in optical cavities,
the light-matter interaction down to the few-photon or vacuum limits 
attracted considerable attention
~\cite{GarciaVidal_Science_2021, Genet_PT_2021, Rubio_NatureMater_2021, Bloch_Nature_2022, 
Schlawin_APR_2022, hubener_cavity_perspective, scalari_science_2012, muravev_prb_2013, maissen_prb_2014, 
smolka_science_2014, Keller_nanolett_2017, Paravicini_Bagliani_NaturePhys_2019, 
appugliese_iqhe2022, jarc_nature2023, 
graphene_plasmonic_cavities_kim2023, mazza_superradiant_2019, andolina_no-go_prb_2019,andolina_condensation_prb_2020, 
mazza_polini_hidden_excitonic_prb2023,
andolina_perturbative_no-go_epjp_2022, amelio_prb, andolina_amperean_superconductivity_2024, 
riolo2024tuningfermiliquidssubwavelength}.

\begin{figure}[t!]
\includegraphics[width=\columnwidth]{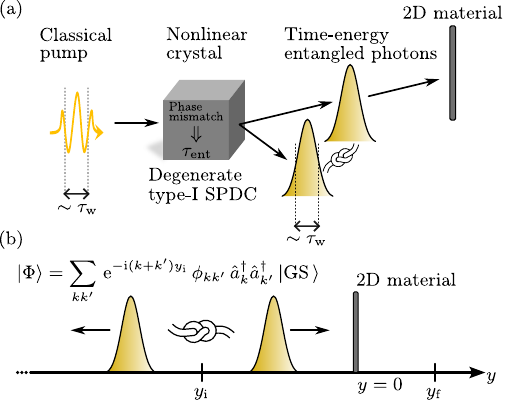}
    \caption{
    Panel (a) Sketch of the generation of time-energy entangled photons. A classical pump of temporal width $\tw$ excites a 
    nonlinear crystal producing two time-energy entangled photons via SPDC. 
    The phase mismatch between the wave-vectors of the classical pulse 
    and the generated pair defines the entanglement time $\tent$.
    The knot represents the entanglement carried by the photon pair.
    One of the two partners of the entangled photon pair scatters onto a 2D material.
    Panel (b) Sketch of our theoretical model. The time-energy entangled pair consists of two counter-propagating photon packets moving along the $y$ direction, see Eq.~\eqref{eq:two-photon_state}. 
    The right-propagating packet
     interacts 
    with a collective excitation hosted by a 2D material
    placed at $y=0$.      
    The symbol
    $\yi$ [$\yf$] indicates the position
   at which the photon pair is created [the transmitted photons are measured].}
    \label{fig:fig1}
\end{figure}

Among the unique quantum properties of light, entanglement 
plays a prominent role for a wide spectrum of applications,
ranging from experimental tests of the foundations of quantum mechanics~\cite{aspect_experimental_1982,ou_violation_1988}
to quantum metrology~\cite{giovannetti_quantum-enhanced_2004, giovannetti_quantum_2006}, imaging~\cite{moreau_imaging_2019},
and the development of quantum technologies~\cite{Slussarenko,Wang_NaturePhoton_2020}.
Advancements in quantum light generation 
stimulated a considerable interest in spectroscopic applications 
of entangled photons~\cite{dorfman_nonlinear_2016, szoke_entangled_2020, mukamel_roadmap_2020}. 
Notable examples include two-photon absorption spectroscopy~\cite{gea-banacloche_two-photon_1989, dayan_two_2004, richter_collective_2011, leon-montiel_temperature-controlled_2019, javanainen_linear_1990},
which is an established tool for a broad range of quantum chemistry applications~\cite{schlawin_two_photon_acs2018,matsuzaki_superresolution_2022, chen_entangled_2022,eshun_ent_photon_spectroscopy_acs2022}.
In many cases, entangled photons are generated via
spontaneous parametric down conversion (SPDC) mechanisms.
These can generate time-energy entangled pairs---see Fig.~\ref{fig:fig1}(a)---characterized by a correlation between 
the frequencies of the two photons~\cite{saleh_virtual_state_1998,yabushita_frequency_entangled_2004,baek_spectral_properties_2008,kalashnikov_infrared_light_2016,arahata_tunable_absorption_2022}. 
Additionally, SPDC-generated pairs can show also polarization entanglement~\cite{shih_typeII_1995,shih_biphoton_field_2003}.
Time-energy entangled photons are particularly useful in combination 
with time-resolved spectroscopy techniques~\cite{dorfman_stimulated_2014,schlawin_pump-probe_2016,
chen_vibrational_2021,zhang_entangled_2022}.
For example, they
have been exploited in pump-probe experiments, to investigate the dynamics of 
excited-states~\cite{ishizaki_probing_2020, fujihashi_probing_2023}, and in interferometric setups to measure femtosecond
dephasing times using continuous-wave pulses~\cite{Kalashnikov_quantum_2017}. 

So far, the theoretical investigation of the effects of the
photon entanglement in the light-matter interaction processes 
focused, almost exclusively, on the properties of light. 
For example, spectroscopic signals involving 
entangled photons have been determined for various setups by means of perturbative methods---see e.g. Ref.~\cite{dorfman_nonlinear_2016} for a review. While relevant predictions have been made, these approaches do not generally provide 
information on the properties of the matter degrees of freedom 
excited by entangled photons.
In this Article, we point out that 
the understanding of the non-equilibrium dynamics of matter stimulated by entangled light 
can be of fundamental interest, greatly transcending applications in the context of quantum spectroscopy.  
In particular, we highlight the fact that matter degrees of freedom 
are naturally characterized by an intrinsic entanglement, which can potentially be controlled, for example, by classical~\cite{Baykusheva_neq_fisher_PRL2023,
Hales_light_driven_entanglement_NATCOMM2023,sun_basov_entangled_plasmons_prr2022} or vacuum fields~\cite{Mazza_PRB_2024,kass_many-body_quantum_blockade_2024}. Therefore, a deep understanding of the hybridization between 
entangled photons and matter degrees of freedom represents 
a fundamental step for opening new frontiers in the exploration 
of light-matter interactions, and, possibly, guiding 
the search for exotic light-induced phases of matter.

In this Article, we study an {\it exactly solvable} model which 
allows us to 
 describe the dynamics of hybrid light-matter excitations 
(i.e.~polaritons~\cite{Basov_Polaritons_2025,basov_polariton_materials_science_2016}) induced by the stimulation 
of matter degrees of freedom by time-energy entangled photons. Specifically, we consider a scattering problem in which 
a single partner of a time-energy entangled photon pair 
interacts with a resonant (non-dispersive) medium described by a  
collective matter excitation hosted by a 2D material, see Fig~\ref{fig:fig1}(b). By varying the time-energy correlations of the incident 
photon, we show that entanglement induces qualitative changes 
in the dynamics of both transmitted photons and light-stimulated matter excitations.

By using incident packets of fixed temporal width, 
we show how entanglement significantly enhances 
transmission across the material. 
The transmission enhancement is characterized by a crossover from delayed to synchronized 
transmission and absorption peaks occurring, respectively,
for weakly and strongly entangled incoming pulses.
Focusing on the dynamics of 
light-stimulated matter excitations, 
we show that  
the entangled driving modifies the time scales of energy transfer 
from the light to the matter degrees of freedom. In particular, we show how energy is absorbed on a 
time scale shorter than the packet width, thus highlighting
how temporally-wide entangled photon packets share some 
similarities with ultrashort light pulses.   
We understand these behaviors by showing 
that an entangled pulse can be represented by 
a superposition of train of pulses containing 
an increasing number of shorter and shorter 
non-entangled packets. Eventually, we show how the 
polariton dynamics leads to 
the creation of matter excitations characterized by 
temporal correlations of time-energy entangled modes 

Our Article is organized as following.
In Section~\ref{sec:model} we introduce our model
and describe the entangled-photon drive considered in this work. 
In Section~\ref{sec:results} we present our main results. 
This Section is organized into three parts. In Sect.~\ref{sec:light_dynamics} 
we discuss the results relative to the dynamics of photon transmission processes through the material and introduce a representation of entangled packets in terms of a superposition of non-entangled train pulses.
Section~\ref{sec:matter_dynamics} contains 
results concerning the dynamics of the light-stimulated matter excitation,
whereas, in Section~\ref{sec:temporal_correlations}, we
discuss the temporal correlations induced by the entangled 
photon driving.
Finally, in Sect.~\ref{sec:conclusions} we summarize our main findings and draw our conclusions.

\section{Setup and entangled photons}
\label{sec:model}
We consider a scattering setup in which a single photon 
propagating along the $y$-axis impinges on a 2D 
material in the $x$-$z$ plane at $y=0$, see Fig.~\ref{fig:fig1}(b). The incident photon is entangled with another photon 
which propagates in the opposite direction and does 
not interact with the material. 

We describe entangled 
photons by populating the modes of the electromagnetic field quantized 
in a cavity of length $L_y$~\cite{kakazu_quantum_cavities}.
We take the cavity large enough so that, 
for all practical purposes, we deal with the continuum of free-space 
electromagnetic modes.
Each photon mode is labeled by an in-plane wave-vector
$\bm{q}$, an out-of-plane wave-number $k$, and an associated 
polarization $\l$. Here, ``out-of-plane'' and ``in-plane'' are referred to the $x$-$z$ plane of the 2D material.

In this Article, we consider the interaction between light and a 
single collective mode of the material with 
zero in-plane wave-vector, $\bm{q} = {\bm 0}$, and polarization along the $x$-axis,
so that our theory includes only photonic modes 
with $\bm{q}={\bm 0}$ and a single polarization $\l=\hat{x}$.
From now on, we drop the dependence on these two indices.
We model the collective mode of the matter degrees of freedom 
as an harmonic oscillator of frequency $\o_0$. 
This mode may represent, for example, a zone-center 
optical phonon, a 2D gapped plasmon~\cite{novelli_PRB_2020,cavicchi_arXiv_2024,abedinpour_prl_2007}, an exciton or a magnon in an atomically-thin semiconductor~\cite{basov_polariton_materials_science_2016,Basov_Polaritons_2025}.
For the sake of concreteness, we fix the energy 
of the collective mode at $\hbar\o_0 = 100~\mathrm{meV}$,
without further specifying its microscopic origin. 

Upon minimal coupling substitution~\cite{grynberg_introduction_2010,loudon_quantum_2000},
the light-matter Hamiltonian therefore reads as following:
\begin{equation}
    \hat{\mathcal{H}}
    =
    \hat{\mathcal{H}}_{\rm mat}
    + \hat{\mathcal{H}}_{\rm pht}
    + \hat{\mathcal{H}}_{\rm mat-pht}\, .
    \label{eq:total_hamiltonian}
\end{equation}
Here,  $\hat{\mathcal{H}}_{\rm mat} = \hbar \o_0 \bc \ba $
and $\hat{\mathcal{H}}_{\rm pht} = \sum_{k} \hbar \o_k \ac_k \aa_k$ with $\o_k = c |k|$ are, respectively, 
the bare matter and photon Hamiltonians, where $\ba$ [$\bc$] and $\aa_k$ [$\ac_k$] 
are the corresponding bosonic annihilation [creation] operators and $c$ is the speed of light in vacuum.
The last term in Eq.~(\ref{eq:total_hamiltonian}), $\hat{\mathcal{H}}_{\rm mat-pht} = \hat{\mathcal{H}}_{A} + \hat{\mathcal{H}}_{A^2}$, represents light-matter interactions and contains paramagnetic and diamagnetic contributions, which read, respectively, 
\begin{subequations}
\begin{equation}
    \hat{\mathcal{H}}_A=
    -\mathrm{i}\hbar \omega_0
    \sum_{k}
    \frac{\tilde{g}}{\sqrt{\tilde{\omega}_{k}}}
    \left(
    \bc \aa_{k} + \bc \ac_k
    \right)+\mathrm{H.c.}\, ,
    \label{eq:paramagnetic_hamiltonian}
\end{equation}
    \begin{equation}
        \hat{\mathcal{H}}_{A^2}=
    \hbar \omega_0
    \sum_{k,k'}
    \frac{\tilde{g}^2}{ \sqrt{\tilde{\omega}_{k}\tilde{\omega}_{k'}} }
    \left(\ac_k \aa_{k'}+
    \ac_{k} \ac_{k'}\right)
    +\text{H.c.}\, .            
    \label{eq:diamagnetic_hamiltonian}
\end{equation}
\end{subequations}
Here, $\tilde{\o}_k \equiv \o_k/\o_0$ is the dimensionless photon frequency (i.e.~the photon frequency $\omega_k$ measured in units of $\omega_0$)
and 
$\tilde{g} \equiv \tilde{q} \sqrt{\alpha \pi c/(L_y \o_0)}$ is a dimensionless coupling 
defined by the fine structure constant $\a= e^2/(\hbar c)\simeq 1/137$ and the dimensionless 
effective charge $\tilde{q}$
of the collective mode. Unless specified elsewhere, 
we set $\tilde{q} = 1$ and introduce a high-energy cut-off $\L= 2\omega_0$ 
on the photonic modes. All the  results presented below are converged 
with respect to this cutoff.

The Hamiltonian~\eqref{eq:total_hamiltonian} is bilinear 
in the bosonic annihilation/creation operators and, upon introducing polariton operators~\cite{Hopfield} $\ga_{a}$ and $\gc_{a}$, can be brought in 
diagonal form  $\hat{\mathcal{H}} = \sum_{a} \varepsilon_a \gc_{a} \ga_{a} $ 
by a multi-mode canonical para-unitary transformation~\cite{colpa_diagonalization_1978}, 
see Appendix~\ref{sec: time evolution}. 
The vacuum state of the polaritonic modes,  $\ga_a$ and $\gc_a$, 
defines the ground state of the Hamiltonian, i.e.~$\ga_a \ket{\rm GS} = 0$.
Notice that, due to the presence of terms of the form $\hat{b}^\dagger\hat{a}_k^\dagger$,  $\hat{a}^\dagger_k\hat{a}^\dagger_{k'}$ and Hermitian conjugates in Eqs.~(\ref{eq:paramagnetic_hamiltonian}) and (\ref{eq:diamagnetic_hamiltonian}), 
the $\ket{\rm GS}$ does not coincide with the 
vacuum of bare photon and matter operators $\aa_k \ket{0} = \ba \ket{0} = 0$.
As a result, the resonance frequency gets dressed 
by the light degrees of freedom  and undergoes a blue shift 
$\o_0 \to \o_* \simeq 1.008\o_0$. Further details about the preparation of the initial state and this blue shift of the matter resonant frequency can be found in Appendix~\ref{sec:initialization} and \ref{sec:spontaneous_decay_rate}, respectively.

Starting from $\ket{\rm GS}$ at $t=0$, we suddenly change the state 
of the system by creating a two-photon Fock state describing  
counter-propagating packets centered at an 
initial position $\yi<0$ on the left side of the material, see Fig.~\ref{fig:fig1}.
The sudden creation of the two-photon state brings the system 
out-of-equilibrium and the unitary dynamics 
governed by the Hamiltonian in Eq.~\eqref{eq:total_hamiltonian} describes the spatio-temporal propagation of the 
two packets and, in particular, the interaction 
of the right-moving packet with the material. 
The $t=0$ two-photon state reads
\begin{equation}
    \ket{\Phi} 
    =
    \sum_{k>0}
    \sum_{k'<0} 
    \mathrm{e}^{-\mathrm{i}(k+k') \yi}
    \phi_{kk'}\,
    \hat{a}^\dagger_{k} \hat{a}^\dagger_{k'}
    \ket{\rm GS}\, ,
    \label{eq:two-photon_state}
\end{equation}
where the frequency-dependent coefficients, 
$\phi_{kk'} = \phi(\o_k,\o_{k'})$,
define the entanglement properties of the two-photon state.
In this Article, we consider a time-energy entangled pair in which 
entanglement is controlled by the correlation between the frequencies of the two 
counter-propagating packets. 
Specifically, we consider a simplified version of a 
type-I SPDC
state~\cite{hong_theory_1985, joobeur_spatiotemporal_1994, law_continuous_2000, 
ma_multimode_1990, loudon_quantum_2000,grynberg_introduction_2010}
defined by two time scales, $\tau_{\rm w}$ 
and $\tent$ as
\begin{equation}
    \begin{split}
        \phi_{kk'} & = {\cal N}_{\rm ent}\,
        {\rm e}^{-(\o_k+\o_{k'}- 2\o_0)^2 \tw^2}\\ 
        & \times \mathrm{sinc}\left(
        (\omega_k-\omega_0)^2 \tent^2
        +(\omega_{k'}-\omega_0)^2 \tent^2\right)\,,
        \label{eq:entangled_state_envelope}
    \end{split}
\end{equation}
where 
${\rm sinc}(x) =\sin (x)/(x) $ and 
${\cal N}_{\rm ent}$ is a normalization factor such that $\langle \Phi|\Phi \rangle=1$.
The two-photon state in Eq.~\eqref{eq:two-photon_state} can be 
generated by stimulating a birefringent material with a classical pulse.
The time scales $\tw$ and $\tent$ are 
related, respectively, to the width of the pulse, $\sigma_{\rm w} \sim \tw^{-1}$, and  the wave-vector mismatch inside the nonlinear crystal~\cite{baek_spectral_properties_2008}.
The quantity $\phi_{k k'}$ gives the probability amplitude 
of measuring two photons with momenta $k$ and $k'$. We notice from Eq.~\eqref{eq:entangled_state_envelope}  that the coefficients $\phi_{k k'}$ 
cannot be factorized, i.e.~$\phi_{k k'} \neq \psi_{k} \psi_{k'}$.
As a result, the state $\ket{\Phi}$ cannot be written as a product state of left- and right-moving packets, i.e.~ 
$\ket{\Phi} \neq \ket{\Phi_{\rm L}} \ket{\Phi_{\rm R}}$, highlighting its entangled nature. 
\begin{figure}[t]
    \includegraphics[width=\columnwidth]{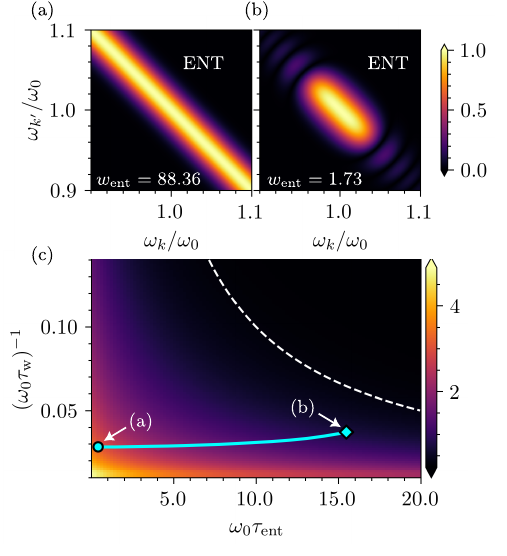}
    \caption{Panels (a)-(b)
    Modulus $|\phi_{kk'}|$ of the mode-dependent coefficients $\phi_{kk'}$,
    see Eq.~\eqref{eq:entangled_state_envelope}, defining the two-photon entangled state in Eq.~\eqref{eq:two-photon_state}. The physical meaning of $\phi_{kk'}$ is explained in the main text.
    The mode-dependent coefficients are normalized to their maximum value.
    Panel (a) Results for $\tw \approx 232.02~{\rm fs}$ and  $\tent \approx 2.63~{\rm fs}$, corresponding 
    to an entanglement parameter $\went \approx 88.36$, see Eq.~\eqref{eq:ent_parameter}.  
    Panel (b) Results for $\tw \approx 176.59~{\rm fs}$ and $\tent \approx 101.88~{\rm fs}$, corresponding to 
    $\went \approx 1.73$. Panel (c) The von Neumann entropy is plotted as a function of 
    $(\omega_0 \tw)^{-1}$ and $\omega_0\tent$. The white dashed line corresponds to $\went=1$, where $\went$ is defined in Eq.~\eqref{eq:ent_parameter}. States carrying strong
    entanglement fall below the dashed line. The cyan line represents the locus of points in parameter 
    space for which the entangled packets have a fixed temporal width corresponding to a FWHM of approximately $556~{\rm fs}$. 
    The dot and diamond symbols denote the points corresponding to panels (a) and (b) above.} \label{fig:fig2}
\end{figure}

The non-separability of the state is 
determined by the frequency correlations set by 
the exponential and sinc factors in Eq.~\eqref{eq:entangled_state_envelope} and controlled by the two time scales $\tent$ and $\tw$. 
In the limit $\tau_{\rm ent} \ll \tw $, the 
frequencies of the two packets are anti-correlated around the 
resonance frequency, i.e.~$\o_{k'} \approx 2 \o_0 - \o_k$, over a broad range 
of frequencies, see Fig.~\ref{fig:fig2}(a). 
By increasing the parameter $\tent$, the 
frequencies become anti-correlated on a much smaller 
frequency range---Fig.~\ref{fig:fig2}(b)---implying that, for $\tent \gg \tw$, the state 
evolves towards a separable one, i.e.~$\ket{\Phi} \approx \ket{\Phi_{\rm L}} \ket{\Phi_{\rm R}}$.
We confirm these expectations by direct calculation 
of the entanglement entropy associated to the partitioning of the 
photon Hilbert space into two sectors, one corresponding to positive value of the wave-number $k$ and one to negative values of $k$,
representing, respectively, right- and left-moving modes, 
see Appendix~\ref{sec:entanglement_entropy} for details.
Panel (c) of Fig.~\ref{fig:fig2} displays 
the entanglement entropy as a function of 
$(\omega_0 \tw)^{-1}$ and $\omega_0\tent$. 
The entanglement entropy is large for $\tw/\tent \gg 1$ while it is strongly 
suppressed for $ \tw/\tent \lesssim 1$. 

We quantify the the temporal widths of the 
packets by computing their full width at half maximum (FWHM). 
For the packets defined by Eq.~\eqref{eq:entangled_state_envelope}, the FWHM 
is controlled by both time scales $\tw$ and $\tent$.
In the following, we have decided to fix the temporal width 
of the entangled packets and change the two time 
scales to control the entanglement entropy of the state.
Specifically, we introduce the parameter
\begin{equation}
    \went \equiv \frac{\tw}{\tent}
    \label{eq:ent_parameter} 
\end{equation} 
and restrict our analysis to the parameter region 
$\went >1$, see cyan line in Fig.~\ref{fig:fig2}(c). 
Strongly [weakly] entangled packets occur for
for $\went \gg 1$ [$\went \gtrsim 1$]. For $\went>1$ it is possible to check that the entangled packets described by Eq.~(\ref{eq:two-photon_state}) with the choice (\ref{eq:entangled_state_envelope})
have a Gaussian shape. 
Therefore, we compare the results obtained with entangled packets, evaluated at different values of $\went$, 
with those obtained for rigorously separable Gaussian packets defined by
\begin{equation}
    \phi^{\rm sep}_{kk'}
    =
    \psi_k \psi_{k'}
    =
    {\cal N}_{\rm sep}
    \mathrm{e}^{-(\omega_{k}-\omega_0)^2 \tsep^2}
    \mathrm{e}^{-(\omega_{k'}-\omega_0)^2 \tsep^2}.
    \label{eq:phikk_sep}
\end{equation}
Here, ${\cal N}_{\rm sep}$ is a normalization constant 
and $\tsep$ defines the temporal width of the separable packets.
For the separable packets, the FWHM depends only on $\tau_{\rm sep}$.

\begin{figure}
    \includegraphics[width=\columnwidth]{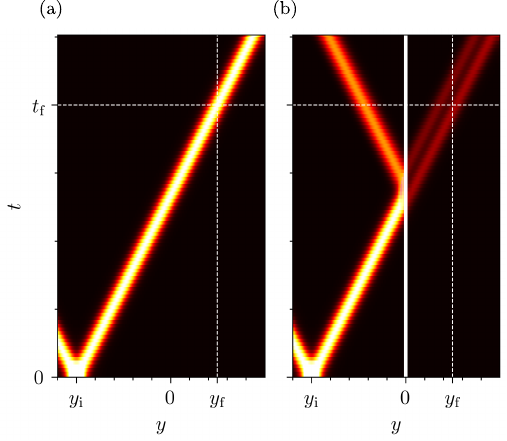}
    \caption{Spatio-temporal evolution of the photon number density $n_{\rm pht}(t,y)$, see Eq.~\eqref{eq:photon_number_density},
    of the left- and right-propagating entangled packets created at time $t=0$ at position $y_{\rm i}$. Numerical results in this figure refer to the case of entangled packets with $\went = 1.73$. 
    Panel (a) Results for the case of freely-propagating packets. Panel (b) The 
    right-propagating packet interacts with the 2D material at $y=0$, here represented by a thick white line. 
    In both panels, the evolution of the left-propagating packet is cut for visualization purposes. 
    Vertical [horizontal] dashed lines represent the position $y_{\rm f}$ where transmitted photons 
    are measured  [the time $t_{\rm f}$ at which the freely propagating packet reaches the position $y_{\rm f}$]. 
    }
    \label{fig:fig2.5}
\end{figure}

\begin{figure}[t]
    \includegraphics[width=\columnwidth]{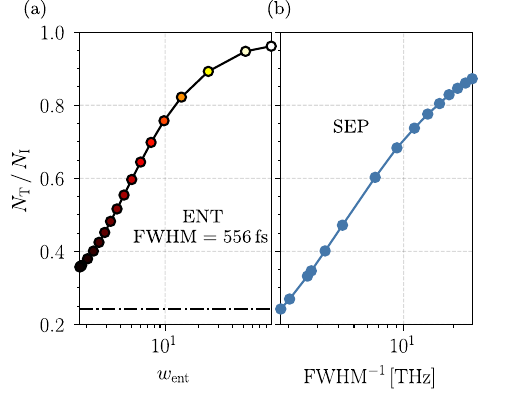}
    \caption{
    Panel (a) Number of transmitted photons for entangled packets at a fixed ${\rm FWHM} \approx 556~{\rm fs}$, 
    as a function of the entanglement parameter $\went$---see Eq.~\eqref{eq:ent_parameter}. 
    The color scale on the dots indicates the value of the entanglement entropy, from minimum (dark) to 
    maximum (bright). The horizontal dash-dotted line indicates the number of transmitted photons for a 
    separable packet of the same FWHM. Panel (b) Number of transmitted photons for separable packets as a 
    function of the inverse of the FWHM. The separable packet with the largest width in panel (b) has the 
    same FWHM as the entangled packets in panel (a). \label{fig:fig3}}
    \end{figure}

\section{Results}
\label{sec:results}
In the following sections we present results obtained by numerically solving 
the equations of motion for the coupled 
photon and matter degrees of freedom.
As already mentioned above, the Hamiltonian 
in Eq.~\eqref{eq:total_hamiltonian} can be 
diagonalized exactly. The corresponding equations of motion therefore describe the exact 
unitary dynamics of the coupled 
light-matter system.

We organize the presentation of our results into three parts.
Sect.~\ref{sec:light_dynamics} and~\ref{sec:matter_dynamics} 
are devoted to time-dependent observables of photonic and matter 
degrees of freedom, respectively. 
Finally, in Sect.~\ref{sec:temporal_correlations}, we discuss non-local temporal correlations of both degrees of freedom.

Before presenting our results, 
it is important to specify some numerical details regarding the system under consideration. 
In the setup shown in Fig.~\ref{fig:fig1}~\added{(b)}, periodic boundary conditions (PBCs) are imposed at $y=-\frac{L_y}{2}$ and $y=\frac{L_y}{2}$. The two photon wave packets are initialized at time $t=0$ in the region $y<0$ and travel in opposite directions. Consequently, the left-moving wave packet eventually crosses the boundary, re-enters from the opposite side, and travels towards the 2D material located at $y=0$. At sufficiently long times, this wave packet will either overlap with the 
right-propagating packet or interact with the 2D material. In the results presented in this work, such artifacts arising from the application of PBCs have been carefully avoided by limiting the simulation time. Thus, when we refer to ``asymptotic'' expectation values for long times (e.g.~$t \to +\infty$, as in the case of 
Sect.~\ref{sec:light_dynamics}), we mean that these times are long enough for the interaction between the right-moving wave packet and the 2D material to be complete, yet short enough to eliminate artifacts due to PBCs.
\subsection{Photon dynamics}
\label{sec:light_dynamics}
\begin{figure*}[t]
    \includegraphics[width=\textwidth]{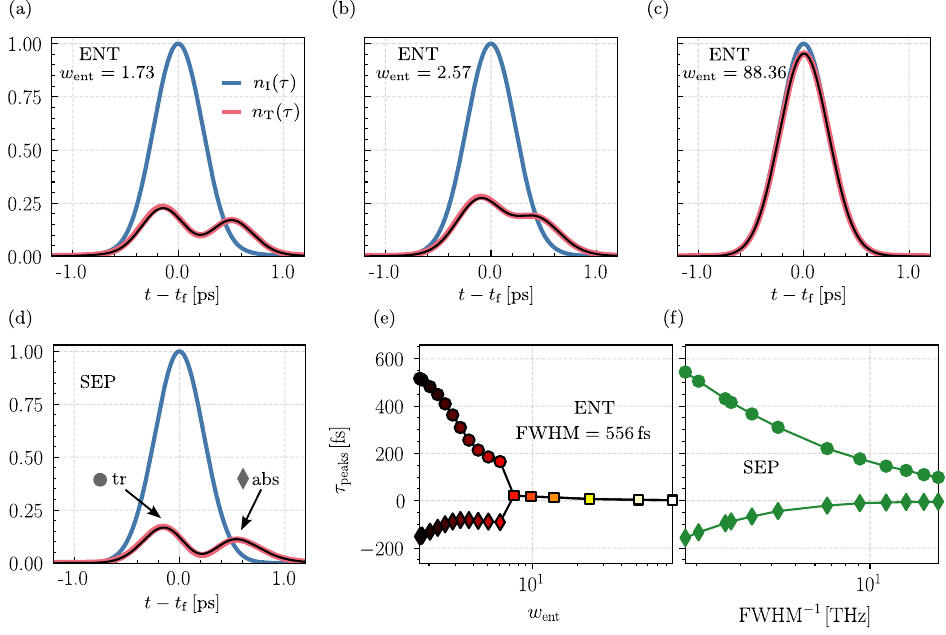}
    \caption{Panels (a)-(d)
    Temporal evolution of the density of transmitted photons (orange line) 
    as a function of the delay time $\t=t-t_\mathrm{f}$, 
    where $t_{\mathrm{f}}$ is the arrival time 
    of the peak of the freely propagating photon wave packet (blue lines) 
    at the same position $y_\mathrm{f}\gg0$. The incident wave packets 
    have a FWHM of approximately $556~{\rm fs}$ in all panels from (a) to (d). 
    All the photon densities are normalized to the peak value of the corresponding incident density. In panels (a)-(c) we display the temporal evolution of entangled packets. In particular, results in panel (a) have been obtained by setting  $\went\approx1.73$ and $\tent \approx101.88~{\rm fs}$. Results in panel (b) correspond to: $\went \approx2.57$ and $\tent \approx79.17~{\rm fs}$. Finally, results in panel (c) correspond to $\went \approx88.36$ and $\tent \approx2.63~{\rm fs}$. Panel (d) refers to a separable packet with $\tsep \approx232.71~{\rm fs}$.
    The entanglement parameter $\went$ and the time scales $\tent$ and $\tsep$ are defined in 
    Eqs.~(\ref{eq:entangled_state_envelope})-(\ref{eq:phikk_sep}).
    The thin black lines represent the transmitted densities 
    evaluated by using Eq.~(\ref{eq:transmitted_photon_number_Tk}).
    The circle and diamond symbols in panel (d)
    highlight the transmission (tr) and absorption (abs) 
    peaks of the transmitted packet.
    Panel (e) Transmission (circles) and absorption (diamonds) 
    peaks for entangled packets with $\mathrm{FWHM}\approx556~{\rm fs}$ as
    functions of the entanglement parameter $w_{\rm ent}$. 
    The squares indicate the single peak of the transmitted 
    packet when the distinction between  transmission 
    and absorption peaks is lost. The color scale on the symbols of panel (e) indicates the value of the entanglement entropy characterizing the incident packet, 
    from minimum (dark) to maximum (bright).
    Panel (f) Transmission (circles) and absorption (diamonds) 
    peaks for separable packets as functions of the inverse FWHM.}
    \label{fig:fig4}
\end{figure*}

We start by looking at the transmission of the photon beam across the 2D material. 
We compute the spatio-temporal evolution of the linear density of 
photons
\begin{equation}
    n_{\mathrm{pht}}(t, y)
    =
    \frac{1}{L_y}
    \sum_{kk'}
    \mathrm{e}^{-\mathrm{i}(k-k')y}\,
    \quave{\ac_{k}\aa_{k'}}(t)\,,
    \label{eq:photon_number_density}
\end{equation}
where $\quave{\ac_{k}\aa_{k'}}(t) \equiv \braket{\Phi}{\ac_{k}(t) \aa_{k'}(t)}{\Phi}$
with $ \ac_{k}(t)  =  e^{i \hat{\cal H} t}  \ac_{k} e^{-i \hat{\cal H} t} $
and $ \aa_{k'}(t)  =  e^{i \hat{\cal H} t}  \aa_{k'} e^{-i \hat{\cal H} t} $.
Spatial integration of the linear density gives the total number of 
photons
\begin{equation}
    N_{\rm pht}(t)
    =
    \int_{-\frac{L_y}{2}}^{\frac{\phantom{-}L_y}{2}} \de y ~n_{\mathrm{pht}}(t, y)
    =
    \sum_{k} \quave{\ac_{k} \aa_{k}}(t)\,.
\end{equation}
It is important to note that, because of the light-matter interaction, 
the total number of photons is not conserved during the dynamics, 
and it reduces to its initial value in the asymptotic 
limit $N_{\rm pht}(t\to \infty) = N_{\rm pht}(0)$.

In Fig.~\ref{fig:fig2.5}, we show two examples of the spatio-temporal evolution 
of the left- and right-propagating entangled packets for $\went =1.73$.
Fig.~\ref{fig:fig2.5}(a) shows the free propagation of the packets in the absence 
of the material, whereas Fig.~\ref{fig:fig2.5}(b) shows the interaction of the right-propagating packet with the material 
placed at $y=0$.
Due to the interaction with the material, 
the photon wave packet gets partially transmitted 
and partially reflected.
We define the total transmission by computing the 
fraction of photons found on the right of the 2D material,
in the asymptotic limit
\begin{equation}
    N_{\rm T} = \int_0^{\frac{L_y}{2}} \de y~n_{\rm pht}(t \to \infty,y)\, .
\end{equation}
In Fig.~\ref{fig:fig3}(a), we show the total 
transmission for a series of entangled pulses 
with increasing entanglement parameter $\went$, 
at a fixed FWHM. 
For increasing entanglement parameter $\went$, 
we observe an overall enhancement of the 
total photon transmission.
This can be understood as a direct consequence of the fact that, 
due to the broader distribution of frequencies centered 
around the $\o_0$ resonance, see Fig.~\ref{fig:fig2}(a),
strongly entangled packets contain an increasing number 
of off-resonant components. 
In Fig.~\ref{fig:fig3}(b), we report the transmission 
for separable packets as a function of the 
inverse of their FWHM.
Overall, we observe a striking similarity 
between the transmission of  entangled packets 
of increasing entanglement parameter $w_{\rm ent}$ and fixed FWHM, 
and the transmission for separable packets of decreasing FWHM. 
In the separable case, the transmission increases by reducing the packet width.
By considering the separable packet of the same width of the 
entangled ones, we see that its transmission is smaller than the one for an entangled packet. 
In particular, in the limit of  weakly entangled packets, 
the transmission seems to show a convergent behavior towards the transmission 
of a separable packet of the same FWHM, even though convergence 
is never reached within the range of parameters we have considered.

In Fig.~\ref{fig:fig4} we show the temporal 
evolution of the density of transmitted photons 
computed at a point $\yf > 0 $ at large distance from 
the material, i.e.~$n_{\rm T}(t) \equiv n_{\rm pht}(t,y=\yf)$---see the vertical dashed line in Fig.~\ref{fig:fig2.5}(b). We compare the transmitted density with the incident one $n_{\rm I}(t)$, which is 
defined as the photon density associated with a packet 
that propagates freely without interacting with the material, corresponding to 
a vertical cut in Fig.~\ref{fig:fig2.5}(a).
For visualization purposes, we change time variable from $t$ 
to $\t \equiv t - \tf$, where $t_{\rm f}$ represents the arrival time of the peak of the freely propagating 
packet at the position $\yf$--- see Fig.~\ref{fig:fig2.5}(a).
For weakly entangled ($w_{\rm ent} = 1.73$) and separable packets, 
see Fig.~\ref{fig:fig4}(a) and~(d), respectively, 
the transmitted packets show a double-peak structure indicating 
a temporal delay between the photon density that is transmitted 
without being absorbed by the material 
(peak at negative $\tau$) and the photons 
that are absorbed and re-emitted (peak at positive $\t$).
Notably, as the entanglement parameter $w_{\rm ent}$ increases, the double peak 
structure progressively disappears, see Fig.~\ref{fig:fig4}(b) and~(c). 
From the latter panel (corresponding to $w_{\rm ent}=88.36$)
we clearly see that, for strongly entangled photons, 
the transmitted packet becomes almost synchronous to the incident one. 
In Fig.~\ref{fig:fig4}(e), we show the crossover between the delayed 
and synchronized transmission regimes by plotting the temporal position 
of the two peaks in the transmitted packet as a function of $w_{\rm ent}$. We see that for $w_{\rm ent}\approx 7$ the two peaks disappear, merging onto a single one. 

The reader may be tempted to interpret the existence of a single peak at large $w_{\rm ent}$---see Fig.~\ref{fig:fig4}(c)---as the result of the decrease in height of the absorption peak, i.e.~the peak occurring for positive values of $\tau$ in Figs.~\ref{fig:fig4}(a), (b), and (d), in favor of the transmission peak. 
In order to test this hypothesis, we analyze in Fig.~\ref{fig:fig4}(f) the dependence of the transmission 
and absorption peaks on the FWHM, for the case of separable packets. 
This analysis is pertinent since, as we have seen in Fig.~\ref{fig:fig3}(a), 
by decreasing the FWHM, the total transmission of separable 
packets increases and approaches the transmission of the entangled 
packets with large $\went$.
As clearly seen in Fig.~4(f), separable packets {\it always} 
display two peaks, even in the case of very small values of the FWHM 
on the order of $\approx 53~{\rm fs}$ for which the 
transmission of entangled and separable packets become equivalent. We conclude that the absence of the two-peak structure for large $w_{\rm ent}$ is a genuine consequence of entanglement
and not merely due to the increase of transmission.

\begin{figure}[t]
    \includegraphics[width=\columnwidth]{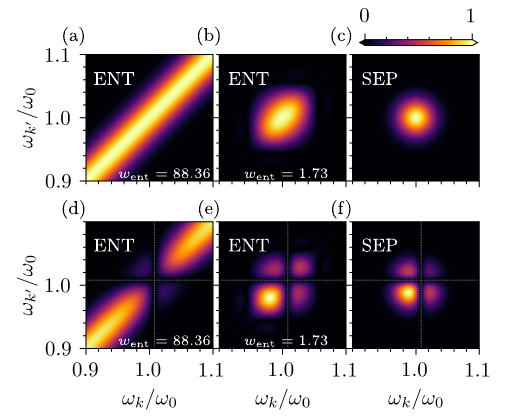}
    \caption{Modulus of the photon number distribution, $|\quave{\ac_{k} \aa_{k'}}|$ for 
    $k,k'\ge0$, of the incident (a)-(c) and transmitted (d)-(f) packets, normalized 
    to their maximum value. Results in panels (a) and (d) refer to the entangled case with $\tw \approx 232.02~{\rm fs}$ 
    and $\tent \approx 2.63~{\rm fs}$, corresponding to $\went \approx 88.36$. Results in panels (b) and (e) refer to the entangled case with $\tw \approx 176.59~{\rm fs}$ and $\tent \approx 101.88~{\rm fs}$, 
    corresponding to $\went \approx 1.73$. Finally, results in panels (c) and (d) refer to the separable case with 
    $\tsep \approx 232.71~{\rm fs}$.
    The entanglement parameter $\went$ and the time scales $\tent$ and $\tsep$ are defined in 
    Eqs.~(\ref{eq:entangled_state_envelope})-(\ref{eq:phikk_sep}).}
    \label{fig:fig5}
\end{figure}

To gain further insight, we investigate the inner structure 
of the density profiles of the transmitted photons. 
From Eq.~(\ref{eq:photon_number_density}), we see that the linear photon density is completely determined by the distribution of 
photon occupation numbers, $\quave{\ac_{k} \aa_{k'}}(t)$, which, 
at a fixed time $t$, depends on the two frequencies $(\o_k,\o_{k'})$.
It is important to stress that nearly identical photon density 
profiles may correspond to very different distributions of photon 
occupation numbers $\quave{\ac_{k} \aa_{k'}}$. 
For example, this is readily seen in panels (a)-(c) of Fig.~\ref{fig:fig5}, where we present plots of $\quave{\ac_{k} \aa_{k'}}_0 \equiv \quave{\ac_{k} \aa_{k'}}(t=0)$ 
for $w_{\rm ent}=88.36$ (strong entanglement), $w_{\rm ent}=1.73$ (weak entanglement), and separable incident packets, respectively. These three situations correspond to the plots in panels (c), (a), and (d) of Fig.~\ref{fig:fig4}, 
where we clearly see that the initial photon density profiles (blue curves) are practically identical.

In panels (d)-(f) of Fig.~\ref{fig:fig5}, we show the same occupation numbers 
$\quave{\ac_{k} \aa_{k'}}(t=t_{\rm f})$ but computed at the instant of time $\tf$, 
that is after the interaction of the wave packets with the 2D material.
The frequency distributions remain identical to the incident 
ones except for a sharp suppression of $\quave{\ac_{k} \aa_{k'}}$ 
along the axes $\o_k=\o_*$ and $\o_{k'} = \o_*$, which
signal the resonant absorption of the components of the initial packets at frequency $\o_*$.
From this observation, we define the 
transmission coefficients $T_k$ by expressing the 
 distribution of photon numbers $\quave{\ac_{k} \aa_{k'}}(t)$ at time $t$ 
in terms of the distribution  
$\quave{\ac_{k} \aa_{k'}}_0$ of the incoming photons.
Specifically, we {\it impose}
\begin{equation}\label{eq:imposition}
    \quave{\ac_{k} \aa_{k'}}(t)
    \stackrel{!}{=}
    |\quave{\ac_{k} \aa_{k'}}_0| T_k^* T_{k'} 
    e^{\im(k-k')(\yi+ct)}\, ,
\end{equation}
where the phase factor takes into account 
the translation of the packet from the initial position
$\yi$, see Fig.~\ref{fig:fig1}. 

We extract the transmission coefficients $T_k$ from Eq.~(\ref{eq:imposition}) by using a single 
broadband separable initial packet and write, for all types of different 
incident packets (i.e.~separable and entangled), the following expression for the transmitted photon density $n_{\rm T}(\tau)$:
\begin{equation}
    n_{\rm T}(\t)
    =
    \frac{1}{L_y}
    \sum_{\{k, k'\}>0}
    \mathrm{e}^{\mathrm{i} (\o_k-\o_{k'}) \t} 
    |\quave{\ac_{k} \aa_{k'}}_0|\, T^*_{k} T_{k'}\, .
    \label{eq:transmitted_photon_number_Tk}
\end{equation}
We follow this procedure to produce the thick black lines in panels (a)-(d) of Fig.~\ref{fig:fig4}. 
We conclude that Eq.~(\ref{eq:transmitted_photon_number_Tk}) 
reproduces extremely well the 
density profiles computed from the exact unitary dynamics. 
The transmission coefficients $T_k$ are well described by the following packet-independent 
expression 
\begin{equation}\label{eq:coefficient_transmission}
    T_{k} = \frac{\o_k-\o_*}{\o_k-\o_* + \im \G/2}\,,
\end{equation}
where $\Gamma$ is the spontaneous decay rate of the matter degrees of freedom, which 
we calculate from perturbation theory, i.e.~$\G\approx6.97\,\mathrm{THz}$---see Appendix~\ref{sec:spontaneous_decay_rate}.

These observations show that all the information about 
the transmitted density $n_{\rm T}(\tau)$ 
is contained in the distribution $\quave{\ac_{k} \aa_{k'}}_0$ of the photon occupation 
numbers of the incident packets.
We therefore propose a simple analytical model 
in which $\quave{\ac_{k} \aa_{k'}}_0$ is parametrized by an 
elliptical distribution in the $(\o_{k},\o_{k'})$ plane, centered around 
the resonance frequency $\o_0$: 
\begin{equation}
    \begin{split}
        |\quave{\ac_{k} \aa_{k'}}_0| &
        \approx 
        N^2 \mathrm{e}^{-[(\o_k-\o_0)^2+(\o_{k'}-\o_0)^2]\frac{\t_1^2+\t_2^2}{4}} \\
        & \times \mathrm{e}^{(\o_k-\o_0)(\o_{k'}-\o_0)\frac{\t_1^2-\t_2^2}{2}}\, .
    \end{split}
    \label{eq:analytical_nkk}	
\end{equation} 
Here, $\t_1$ and $\t_2$ represent the effective major and minor 
semi-axes of the elliptical distribution 
and $N$ is a normalization constant.
For $\t_1 = \t_2$, the distribution is that of a separable packet, 
namely $|\quave{\ac_{k} \aa_{k'}}_0|^2 = n_{k} n_{k'}$, where $n_k$ is the photon occupation number at wave-number $k$.
The Gaussian shape leads to a spherical-symmetric distribution, 
reproducing extremely well  the photon number distribution of 
the separable case shown in panel (c) of Fig.~\ref{fig:fig5}. 
On the contrary, entangled packets are described 
by the choice $\t_2 < \t_1$, which shrinks and enlarges the frequency distribution along 
the  $ \o_{k'} = 2 \o_0 - \o_k$ and $ \o_{k'} = \o_k$ 
axes, respectively, see panels (a)-(b) of Fig.~\ref{fig:fig5}.

Thanks to this parametrization, 
we can transform the momentum sum in Eq.~(\ref{eq:transmitted_photon_number_Tk}) into a frequency integral. Implementing the change $\o-\o_0\to\o,\, \o'-\o_0\to\o'$ of integration variables,
we can rewrite the transmitted photon density as 
\begin{equation}
    \begin{split}
        n_{\rm T}(\t) 
        &= N^2 \iint_{-\infty}^{+\infty}~\de\o~\de\o'
    	\mathrm{e}^{\text{i}(\o-\o')\t}
    	\\
    	&\times
    	\mathrm{e}^{-\o^2\t_+^2/2}\,
    	\mathrm{e}^{-\o'^2\t_+^2/2}
    	\mathrm{e}^{\o\o'\t_-^2}\,
    	T^*(\o)T(\o')\, ,
    \end{split}
\label{eq:number_density_integral_form}
\end{equation}
where we have introduced $2 \t^2_{\pm} \equiv \t_1^2 \pm \t_2^2$. 
Notice that $\t_-$ is always a real variable as $\t_2 < \t_1$. 

Separable packets can be obtained from Eq.~(\ref{eq:number_density_integral_form}) by taking the  $\t_- \to 0$ limit.
In this limit, the integrand in Eq.~(\ref{eq:number_density_integral_form}) factorizes and the 
transmitted photon density can be simply written as the square modulus of a single wavefunction $\Psi_0(\t)$:
\begin{equation}
    \lim_{\t_- \to 0} n_{\rm T}(\t) = \Psi_0^*({\t}) \Psi_0(\t)\, .
\label{eq:elementary0}
\end{equation}
$\Psi_0(\t)$ can be expanded in frequency domain as
\begin{equation}
\Psi_0(\t) = \int_{-\infty}^{+\infty} \mathrm{d}\o\,
\mathrm{e}^{-\text{i}\o\t}\, \psi(\o)\, ,\label{eq:spectral_representatio_Psi}
\end{equation}
where 
\begin{equation}
\psi(\o) = N \mathrm{e}^{-\o^2\t_+^2/2}T(\o)
\label{eq:psielementary}
\end{equation}
and $N$ is a normalization constant.

For entangled packets, i.e.~for a generic value of $\t_-$, we 
expand the exponential factor $e^{\o\o'\t_-^2}$ in Eq.~(\ref{eq:number_density_integral_form}) in powers of $\t_-$ around $\t_-=0$, 
and write the transmitted photon density as
an infinite sum of densities 
associated to elementary wavefunctions $\Psi_j(\tau)$:
\begin{equation}
  n_{\rm T}(\t) = \sum_{j=0}^{\infty} \Psi_j^*(\t) \Psi_j(\t)\, ,
\label{eq:photon_number_expansion}
\end{equation}
where $\Psi_j(\tau)$ is defined by the $j$-th moment
of spectral function $\psi(\omega)$ introduced in Eq.~\eqref{eq:spectral_representatio_Psi}. Specifically, we have
\begin{equation}
\Psi_j(\t) = \frac{\t^j_-}{\sqrt{j!}}  \llangle{\o^j}\rrangle(\t)\,,
\label{eq:pisj_omegaj}
\end{equation}
with
\begin{equation}
\llangle{\o^j}\rrangle(\t)
\equiv \int_{-\infty}^{+\infty} \mathrm{d}\o\, \o^{j}
    \mathrm{e}^{-\text{i}\o\t} \psi(\o)\, .
    \label{eq:j-th_momentum}
\end{equation}
Further technical details are reported in Appendix~\ref{sec:analytical_model}.

Eqs.~\eqref{eq:photon_number_expansion}-\eqref{eq:j-th_momentum} show 
that the transmitted photon density $n_{\rm T}(\tau)$ for entangled packets can be represented as 
an infinite superposition of {\it separable} packets, each characterized by an elementary density $n_j(\t)$ determined by
a single wavefunction $\Psi_j(\t)$,  i.e.~namely
\begin{equation}\label{eq:superposition_trains}
n_{\rm T}(\t) = \sum_{j=0}^{\infty} n_j(\t)\, ,
\end{equation}
with
\begin{equation}
n_j(\t) \equiv \Psi^*_j(\t) \Psi_j(\t)\, .
\label{eq:nj_transmitted}
\end{equation}

As we have seen above, the transmitted photon density in the case of separable packets is represented by a single elementary contribution, see Eq.~\eqref{eq:elementary0}. In contrast, entangled packets, corresponding to $\t_- > 0$, are represented by a 
superposition of non-entangled packets $n_j(\t)$, as in Eqs.~(\ref{eq:superposition_trains})-(\ref{eq:nj_transmitted}),  
with a spread controlled by the parameter $\t_-$, see Eq.~\eqref{eq:pisj_omegaj}. The larger the entanglement parameter, the larger is the number of elementary contributions $n_j(\t)$ that needs to be taken into account to have a faithful representation of the transmitted density.

\begin{figure}[t]
\includegraphics[width=\columnwidth]{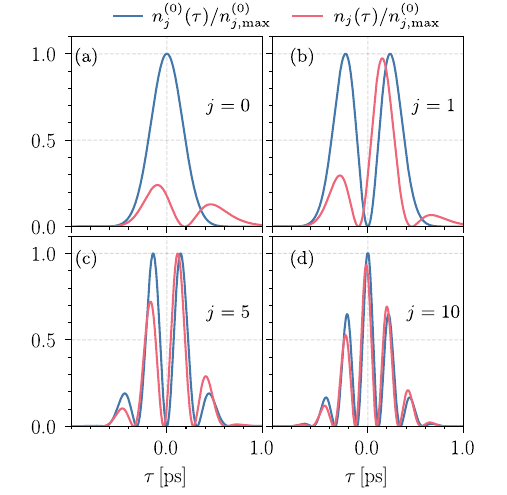}
    \caption{The elementary contribution $n_{j}(\t)$---defined in 
     Eq.~(\ref{eq:nj_transmitted})---plotted as 
    a function of the delay time $\t=t-t_\mathrm{f}$ for different values of $j$. Here, $t_\mathrm{f}$ is the arrival time of the peak of the incident wave packet. 
    Each curve is normalized to the  maximum value $n^{(0)}_{j,\mathrm{max}}$ of the incident packet.  Blue lines represent the incident wave 
    packet---defined in Eq.~(\ref{eq:number_density_analytical_free})---while red lines represent the transmitted packet.
    Data in this figure have been obtained by setting $\t_+\approx232.76~{\rm fs}$ and $\t_-\approx232.67~{\rm fs}$, corresponding to $\t_1\approx329.11~{\rm fs}$ and $\t_2\approx6.58~{\rm fs}$ in Eq.~\eqref{eq:analytical_nkk}. }
    \label{fig:fig6} 
\end{figure}

To further investigate the structure of the elementary contributions, 
we first analyze  $n_j(\t)$ by assuming that no interaction occurs between the incident 
photons and the 2D material. This is achieved by setting 
$\G \to 0$ in the definition \eqref{eq:coefficient_transmission} of the transmission coefficient $T_k$.
In this limit, $T(\o) \to 1$ for all frequencies, and 
the moments $\llangle \o^j \rrangle $ in Eq.~\eqref{eq:j-th_momentum} can be computed analytically. Indeed, the quantity $\psi(\omega)$ in Eq.~\eqref{eq:psielementary} reduces to $\psi^{(0)}(\o) = N \mathrm{e}^{-\o^2\t_+^2/2}$ and the corresponding elementary photon densities can be expressed 
  in closed form as
\begin{equation}
	n^{(0)}_{j}(\t)
	=
	\frac{N^2}{j!}\Big(\frac{\t_-^2}{2\t_+^2}\Big)^j\,
	\mathrm{e}^{-\t^2/\t_+^2}\,
	\mathrm{H}^2_j\left(\frac{\tau}{\sqrt{2}\t_+}\right)\, ,
	\label{eq:number_density_analytical_free}
\end{equation}
where the superscript $(0)$ indicates 
that Eq.~(\ref{eq:number_density_analytical_free}) 
refers to freely propagating photons  and 
$\mathrm{H}_j(x)= (-1)^j e^{x^2} d^j [e^{-x^2}]/dx^j$ are the Hermite polynomials. 
Since Eq.~(\ref{eq:number_density_analytical_free}) displays the product between a Gaussian and the Hermite polynomial $\mathrm{H}_j(x)$, and since the latter has $j$ zeros, the photon density $n^{(0)}_{j}(\t)$---associated with 
each of the separable contributions to the entangled pulse---represents a train of $j+1$ pulses whose width decreases with 
increasing order $j$. 
We explicitly show this in Fig.~\ref{fig:fig6} (blue 
lines) where we plot $n^{(0)}_{j}(\t)$
for different values of $j$.

We now analyze the case of photons interacting with the material by computing the moments $\llangle \o^j \rrangle$ in Eq.~\eqref{eq:j-th_momentum} corresponding to a transmission coefficient $T(\omega)\neq 1$. 
For the sake of analytical amenability, we compute the moments $\llangle \o^j \rrangle$ for $T(\omega)\neq 1$ by taking $\omega_\ast = \omega_0$ in Eq.~(\ref{eq:coefficient_transmission}) 
and carrying out the quadrature in Eq.~(\ref{eq:spectral_representatio_Psi}) analytically, 
see Appendix~\ref{sec:analytical_model}. 
The result of this calculation for different values of $j$ and specific values of $\tau_+$ and $\tau_-$ is shown in Fig.~\ref{fig:fig6} (red curves). 
As seen by comparing incident (blue curves) and transmitted (red curves) photon densities, 
the largest [smallest] contribution to the absorption [transmission] comes from small 
values of $j$.
For these values of $j$, the interaction of photons with the 
collective mode of the 2D material clearly splits the separable 
packets $n_{j}(\t)$ into transmitted and 
absorbed contributions, giving rise to an
additional peak in the density (with respect to the incident density): see red curves in panels (a)-(b) of Fig.~\ref{fig:fig6} corresponding to $j=0$ and $j=1$, respectively.

\begin{figure}[t]
    \includegraphics[width=\columnwidth]{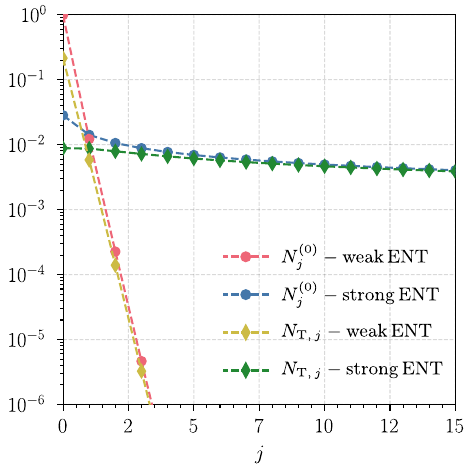}
    \caption{Total number of photons carried by the $j$-th elementary contribution $n_j$
    for the incident wave packets ($N_j^{(0)}$, filled circles) and the transmitted wave packets ($N_{\mathrm{T},j}$, diamonds) in 
    the analytical model. The quantities  $N_j^{(0)}$ and $N_{\mathrm{T},j}$ have been defined in Eqs.~(\ref{eq:Njzero}) and (\ref{eq:Nj_nonzero}), respectively. Results for the strongly entangled [weakly entangled]
    case have been evaluated by setting $\t_+\approx232.76~{\rm fs}$ and $\t_-\approx232.67~{\rm fs}$ [$\t_+\approx325.12~{\rm fs}$ and $\t_-\approx51.08~{\rm fs}$].}
    \label{fig:fig7}
\end{figure}

For $\tau_-=0$ (corresponding to the case in which incident photons are in a separable state), only the $j=0$ elementary contribution 
is different from zero. Our simple analytical model therefore  
consistently captures the double peak 
structure of the transmitted density seen in Fig.~\ref{fig:fig4}(d).
On the contrary, the case of incident entangled photons is described in our analytical model in terms of a large  
superposition of higher-order elementary contributions with $j > 0$.
In particular, for strongly entangled packets, the interaction 
with the collective mode involves the absorption of several elementary 
contributions with $j$ up to $\sim 10$---see panels (c)-(d) in Fig.~\ref{fig:fig6}.
To better appreciate this point, we plot in Fig.~\ref{fig:fig7} 
the total number of photons $N_j$ carried by each $n_j(\tau)$. Specifically, we introduce two quantities:
\begin{equation}\label{eq:Njzero}
N^{(0)}_j \equiv c\int_{-\infty}^{+\infty} {\rm d} \tau\, n^{(0)}_j(\tau)
\end{equation}
and
\begin{equation}\label{eq:Nj_nonzero}
N_{{\rm T}, j} \equiv c\int_{-\infty}^{+\infty} {\rm d}\tau\, n_j(\tau)\, .
\end{equation}

For weakly entangled packets, the number of photons carried by 
each contribution exponentially decays with the order $j$.
As the entanglement parameter increases, the distribution 
evolves towards a power-law decay, showing 
the increasing importance of elementary contributions with large $j$, i.e.~the importance of taking into account multiple separable trains with an increasing number of shorter and shorter pulses. In summary, for strongly entangled packets, as a result of the importance of the $j>0$ contributions in the expansion (\ref{eq:photon_number_expansion}), the interaction process 
between matter and time-energy entangled photons effectively 
includes absorption and re-emission processes taking place on time 
scales much shorter than the temporal width of the incident packet.

We can use these results to rationalize the 
origin of the synchronization phenomenon observed in Fig.~\ref{fig:fig4}.  For large values of $w_{\rm ent}$, Fig.~\ref{fig:fig4}(c), the role of elementary contributions $n_j(\t)$ with large $j>0$ is dominant. Since these are transmitted with no delay---as seen in Fig.~\ref{fig:fig6}(c) and~(d)---the entangled packet in Fig.~\ref{fig:fig4}(c)
gets transmitted as a whole with negligible delay between transmission 
and absorption peaks.

\subsection{Ultrafast light-induced matter excitation}
\label{sec:matter_dynamics}
 \begin{figure}[t]
    \includegraphics[width=\columnwidth]{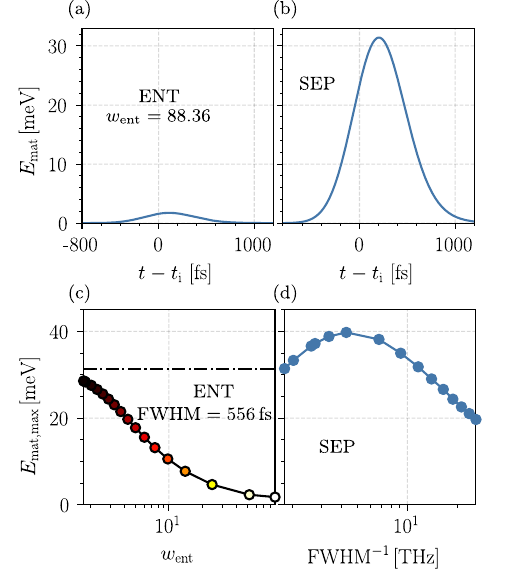}
    \caption{(a)-(b) Energy of the matter degrees of freedom, Eq.~(\ref{eq:energy_matter}), as a function of time measured with 
    respect to the arrival time $t_\mathbf{i}$ of the peak of the incident packet. Panel (a) refers to an entangled packet 
    with $\tent\approx2.63~{\rm fs}$ and $\tw \approx 232.02~{\rm fs}$, corresponding to 
    an entanglement parameter $\went \approx 88.36$. 
    Panel (b) refers to a separable packet for $\tsep \approx 232.71~{\rm fs}$.
    Panel (c) The peak energy induced by entangled packets is plotted as a function of the entanglement parameter $\went$. As usual, the color scale on the symbols in panel (c) indicates the value of the entanglement 
    entropy characterizing the incident packet, from minimum (dark) to maximum (bright). The FWHM of the entangled wave packets in panel (c) has a fixed value of approximately $556~{\rm fs}$. Panel (d) The peak energy induced by separable packets is plotted as a function of the inverse of the
    FWHM. The horizontal
    dash-dotted line indicates the peak energy for a separable packet of the same FWHM of the entangled packets. }
    \label{fig:fig8}
\end{figure}
We now focus on the consequences of the entangled photon driving 
on the dynamics of the light-stimulated matter excitations.  We investigate 
the energy stored in the matter degrees of freedom:
\begin{equation}
    E_{\rm mat} (t) = \hbar \o_0 \langle\Phi|\hat{b}^\dagger(t)\hat{b}(t)|\Phi\rangle \equiv \hbar \o_0 \quave{\bc \ba}(t) \, .
    \label{eq:energy_matter}
\end{equation}
This quantity measures the number of collective excitations induced 
by light in the material.
Experimental characterizations of this quantity usually rely on 
pump-probe setups in which the light-stimulated  material is probed 
after a time delay.
Relevant observables depend on the microscopic details of the excitation 
described by the stimulated collective mode.
Specific examples include effective non-equilibrium 
temperatures determined via transient reflectivity in the case 
of collective excitations of electronic origin~\cite{dalconte_disentangling_electronic_phononic,Brida_NatCommun_2013,Tomadin_PRB_2013,giacomo_temperonics2021} and transient lattice displacements in the case of light-stimulation 
of phonons~\cite{Mankowsky_nonlinear_lattice_for_supercon_nature_2014,nonlinear_phononics}.

In Fig.~\ref{fig:fig8}(a)-(b), we illustrate the 
dynamics of the matter energy for entangled 
and separable incident packets of the same 
temporal width.
Here, we measure time with respect to the 
reference time $\ti \equiv |\yi|/c$ which 
corresponds to the arrival time of 
the peak of the incident photon packet 
at the position of the 2D material (i.e.~$y=0$).
As a function of time, the energy shows a bell-like shape 
indicating power absorption $\partial_t E_{\rm mat} > 0$ followed 
by power emission $\partial_t E_{\rm mat} < 0$.
As a direct consequence of the enhanced transmission
discussed in Figs.~\ref{fig:fig3}-\ref{fig:fig4}, the 
amount of electromagnetic energy absorbed by the 2D material in the entangled case, as measured by the peak value $E_{{\rm mat}, {\rm max}}$
of $E_{\rm mat}$, is much smaller with respect 
to the energy absorbed in the case of a separable 
packet of the same FWHM. In panels (c)-(d) of Fig.~\ref{fig:fig8}, 
we report the peak energies $E_{{\rm mat}, {\rm max}}$ for 
various entangled and separable incident 
packets. 
In the entangled case, Fig.~\ref{fig:fig8}(c), we plot $E_{{\rm mat}, {\rm max}}$
as a function of the entanglement parameter $w_{\rm ent}$, at a fixed FWHM. In the separable case instead, Fig.~\ref{fig:fig8}(d), we plot 
$E_{{\rm mat}, {\rm max}}$ as a function the inverse of the FWHM. These two plots emulate, in spirit, Fig.~\ref{fig:fig3}.

We start by describing $E_{{\rm mat}, {\rm max}}$ for separable packets, Fig.~\ref{fig:fig8}(d). For ultrashort separable packets, 
i.e.~large values of FWHM$^{-1}$, $E_{{\rm mat}, {\rm max}}$ is an increasing 
function of the pulse width 
(decreasing function of FWHM$^{-1}$). 
$E_{{\rm mat}, {\rm max}}$ reaches a maximum for 
FMHW $\approx 191.89~ {\rm fs}$ of the order of the 
inverse decay rate $\sim \G^{-1}$. 
This is due to the fact that, for larger FWHM,
the bandwidth of the incoming photons 
starts to display a smaller overlap with the absorption 
bandwidth of the medium. The increase of $E_{{\rm mat}, {\rm max}}$ with increasing FWHM for short pulses 
is a direct consequence of the decrease
in the number of transmitted photons discussed 
in Fig.~\ref{fig:fig3}(b).
On the other hand, the decrease of $E_{{\rm mat}, {\rm max}}$
for larger values of the FWHM pulses indicates the approach 
towards an ``adiabatic regime'' in which the peak of the energy absorbed by the 2D material  follows the peak of the energy of the incoming pulse.
Moving to the case of entangled packets of 
fixed FWHM, Fig.~\ref{fig:fig8}(c), we observe 
that the peak energy monotonously 
decreases by increasing the entanglement 
parameter $w_{\rm ent}$. In the limit of weak entanglement, $E_{{\rm mat}, {\rm max}}$ 
converges towards the value obtained for a separable packet of the same FWHM.

\begin{figure}[t]
    \includegraphics[width=\columnwidth]{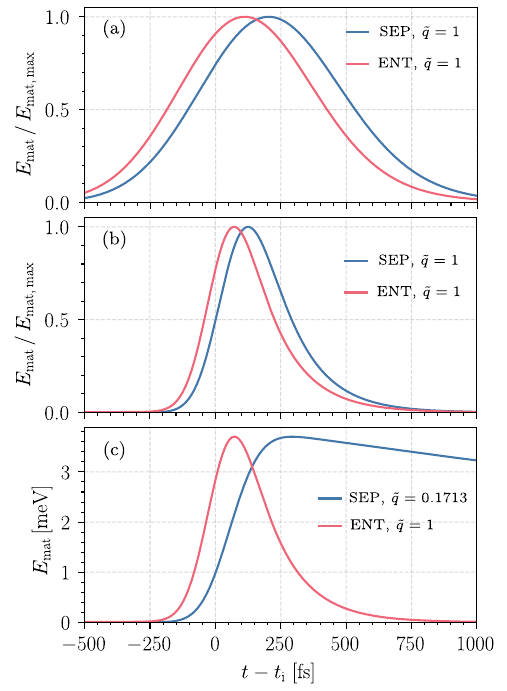}
    \caption{The energy of the matter degrees of freedom, $E_{\rm mat}(t)$, as defined in Eq.~(\ref{eq:energy_matter}), is plotted as a function of time measured 
    with respect to the arrival time $t_\mathrm{i}$ of the peak of the incident wave packet.
    Panel (a) Energy normalized by its peak value induced by an entangled wave 
    packet (orange line) with $\tent\approx2.63~{\rm fs}$ 
    and $\tw\approx232.02~{\rm fs}$, see Eq.~(\ref{eq:entangled_state_envelope}), corresponding to $\went \approx 88.36$, 
    Eq.~(\ref{eq:ent_parameter}), and a 
    separable packet (blue line) with $\tsep \approx 232.71~{\rm fs}$, see Eq.~(\ref{eq:phikk_sep}). (b) Energy
    normalized by its peak value, induced by an entangled wave packet 
    (orange line) with $\tent\approx2.63~{\rm fs}$ and $\tw\approx77.57~{\rm fs}$ corresponding 
    to $\went \approx 29.54$, and a separable packet (blue line) with 
    $\tsep \approx 77.57~{\rm fs}$. (a)-(b) are both evaluated considering an effective charge $\tilde{q}=1$. 
    (c) Energy induced by an entangled wave packet 
    (orange line), evaluated with an effective charge $\tilde{q}=1$, having $\tent\approx2.63~{\rm fs}$, $\tw\approx77.57~{\rm fs}$ 
    and $\went \approx 29.54$, and a separable packet (blue line) with 
    $\tsep \approx 77.57~{\rm fs}$ evaluated with an effective charge $\tilde{q}=0.1713$ such 
    that it matches the matter energy peak value induced by the entangled wave packet.}
    \label{fig:fig9}
\end{figure}

Besides the above quantitative aspects, we now show how entanglement 
in the incoming photon packet qualitatively changes the dynamics of the 
matter degrees of freedom by suppressing the delay between the arrival of the perturbation and the 
creation of the excitation.
In Fig.~\ref{fig:fig9}(a), we show plots of $E_{\rm mat}(t)$ as a function of time, normalized to the corresponding peak values.
In both entangled and separable cases, the peak energy is reached for an 
instant of time $t_{\rm peak} > \ti$ simply indicating 
that, due to causality, energy absorption is 
delayed with respect to the arrival of the perturbation.
However, we clearly see that the peak in the case of the entangled packets 
occurs before the one in the case of separable packets indicating that, 
in the entangled case, energy is absorbed by matter at a faster pace.

In order to confirm that this feature (i.e.~the fact that energy is absorbed at a faster pace in the entangled case) is a genuine consequence of entanglement, we check how the delay 
between the separable and entangled cases depends on 
(i) the width of the packets and (ii) the overall  
absorbed energy. In Fig.~\ref{fig:fig9}(b), we show that such delay persists in the case 
of separable and entangled packets of reduced FWHM.
We notice that, in this case, the temporal profiles $E_{\rm mat}(t)$
acquire an asymmetric shape due to the fact that the FWHM 
of the considered initial photon packets is smaller than the 
spontaneous decay time of the matter degrees of freedom.
We now proceed to rule out the possibility that 
the delay is due to the fact that, for incident 
packets that have the same width, much 
less energy is absorbed in the entangled case 
with respect to the separable one ---as observed in Fig.~\ref{fig:fig8}(a)-(b). To this end, in Fig.~\ref{fig:fig9}(c),
we compare the temporal 
profiles $E_{\rm mat}(t)$ in the case of entangled and separable initial packets of equal widths, and 
giving rise to the same peak energies in the matter degrees of freedom. Specifically, we take two incident packets 
of the same FWHM and, for the separable case, 
we reduce the light-matter coupling constant 
by reducing the effective charge $\tilde{q}$ to make sure that the peak energy in the separable case matches the one reached in the entangled case for $\tilde{q}=1$. The reduced coupling constant in the separable case ($\tilde{q}=0.1713$) suppresses the 
spontaneous decay rate, giving rise to a much more pronounced 
asymmetry between creation and decay of the excitation.
Nonetheless, as we clearly see in Fig.~\ref{fig:fig9}(c), $t_{\rm peak}$ in the case of a separable packet (blue curve) occurs {\it after} the one in the entangled case (red curve).
\begin{figure}[t]
\includegraphics[width=\columnwidth]{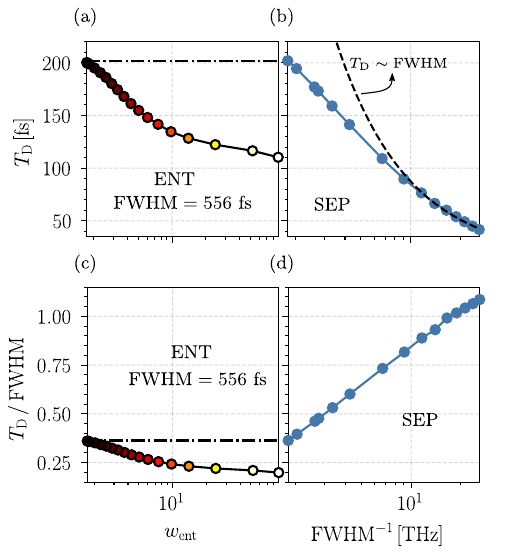}
\caption{Panel (a) The excitation delay time $T_{\rm D}$, defined in the main text as $T_{\rm D} \equiv t_{\rm peak} - t_{\rm i}>0$, is plotted as a function of the entanglement 
        parameter $\went$ for incident entangled wave packets. Panel (b) The delay time $T_{\rm D}$ is plotted as a function of the inverse 
        of the FWHM for incident separable packets. The dashed black line in panel (b) represents the linear trend obtained by fitting the data for the six shortest pulses  with the linear model $T_{\rm D} = {\rm coefficient} \times {\rm FWHM}+{\rm const}$.
        Panels (c)-(d) show the ratio $T_{\rm D}/{\rm FWHM}$. The color scale on the 
        symbols of panels (a) and (c) indicates the entanglement entropy, from minimum (dark) to maximum (bright).
        The horizontal
    dash-dotted lines in panels (a) and (c) indicate $T_{\rm D}$ and $T_{\rm D}/{\rm FWHM}$ for separable packets of the same FWHM 
    of the entangled packets. }
	\label{fig:fig10}
\end{figure}

We now introduce the excitation delay time, $T_D$, defined 
as the difference the time at which the matter energy $E_{\rm mat}(t)$ reaches its maximum value and the time at which the peak in the incident photon density reaches the 2D material, i.e.~$T_{\rm D} \equiv t_{\rm peak} - t_{\rm i}>0$.
Fig.~\ref{fig:fig10} shows the excitation 
delay time $T_{\rm D}$ for separable and entangled incident 
packets, at $\tilde{q}=1$.
For separable packets, panel (b), we plot $T_{\rm D}$ 
as a function of the inverse of the FWHM,  whereas, in panel (a), we plot $T_{\rm D}$ for entangled packets of a fixed FWHM, as a function of the entanglement parameter $w_{\rm ent}$.
For ultrashort separable packets---Fig.~\ref{fig:fig10}(b)---the 
excitation delay time increases linearly with the FMHW.
For wider pulses (smaller FMHW$^{-1}$), the linear trend 
turns into a sublinear growth.  
This can be readily  appreciated by looking at the dashed line   
in Fig.~\ref{fig:fig10}(b) obtained by fitting the data corresponding to the six shortest pulses (i.e.~the six rightmost scattered points in the figure) with a linear model $T_{\rm D} = {\rm coefficient} \times {\rm FWHM}+{\rm const}$.
For entangled packets at fixed FWHM---Fig.~\ref{fig:fig10}(b)---the 
excitation delay converges, for small values of the entanglement parameter,
$\went$ to the value obtained for a separable packet  of the same FWHM.
Remarkably, by increasing $w_{\rm ent}$, $T_{\rm D}$ quickly decreases becoming 
significantly smaller than the FWHM ($556~{\rm fs}$).

Comparing  Fig.~\ref{fig:fig10}(a) with Fig.~\ref{fig:fig10}(b), 
together with the analogous observations in the case of $E_{\rm mat, max}$ in Fig.~\ref{fig:fig8}(c)-(d), we see that, as the entanglement parameter 
increases (for a fixed value of the FWHM), it is as if the matter degrees 
of freedom are being stimulated by separable 
packets of progressively smaller and smaller width.

We notice that the decrease 
of the excitation delay $T_D$ for entangled packets of increasing 
entanglement parameter $\went$ has a sharply different origin, 
as compared to the separable case.
By considering the definition of the entanglement parameter 
$\went$, Eq.~\eqref{eq:ent_parameter}, we observe that the 
increase of $\went$ is controlled by the decrease of $\tent$ 
which is an intrinsic time scale of the entangled pair 
previously introduced in Eq.~(\ref{eq:entangled_state_envelope}).
Specifically, upon reducing $\tent$ the range over 
which the frequencies of the two partners of the entangled pair 
are anti-correlated increases.
This shows that, in the entangled case, the decrease of $T_D$ 
is intimately related to the strong frequency correlation between 
the packet interacting with the material and the packet freely 
propagating in the opposite direction depicted in Fig.~\ref{fig:fig1}.
In contrast, the decrease of $T_{\rm D}$ in the separable case is trivially controlled by the shorter duration of 
the pulse.

To further highlight this difference, we plot in panels (c)-(d) 
of Fig.~\ref{fig:fig10}
the excitation delay $T_{\rm D}$ normalized to the FWHM of the incoming packet.
For separable pulses---panel (d)---$T_{\rm D}/{\rm FWHM}$ increases by decreasing the FWHM. On the contrary, for entangled packets---panel (c)---$T_{\rm D}/{\rm FWHM}$ decreases as a function of the entanglement parameter. We conclude that entangled photons offer a key advantage with respect to separable packets. In the former case,  working with packets of a fixed FWHM, one can induce an ultrafast dynamics of the matter degrees of freedom by increasing $w_{\rm ent}$, i.e.~reducing the inner time scale $\t_{\rm ent}$.

Once again, we can rationalize the above observations 
by invoking the expansion (\ref{eq:photon_number_expansion}) of an entangled packet 
in terms of an infinite superposition of trains of short separable pulses. 
The decrease of the excitation delay $T_{\rm D}$ with increasing $w_{\rm ent}$ is a clear 
signature of the fact that, as shown in Fig.~\ref{fig:fig7}, strongly entangled packets contain trains of ultrafast  separable pulses which are mathematically represented by the elementary 
contributions with $j\gg 1$ in Eq.~(\ref{eq:photon_number_expansion}). As a result, stimulating matter degrees of freedom with entangled packets is similar to stimulating them 
with ultrafast separable packets. At the same time, due to the fact that a large contribution 
of the absorbed energy comes from the $j=0$ elementary contribution, the delay $T_{\rm D}$ in the limit $\went \gg 1$ remains
always larger as compared to 
the ultrashort separable packets, 
as seen in Fig.~\ref{fig:fig10}(a)-(b). 

\subsection{Light-induced temporal correlations}
\label{sec:temporal_correlations}

Having described the ultrafast nature of the 
dynamics of the interaction between matter and entangled photons, 
we now focus on the signatures of the hybridization between 
the two degrees of freedom. 
We do so by investigating the 
temporal correlations 
of the light-stimulated matter degrees of freedom.
Temporal correlations carry information about 
the probability amplitude for the propagation of an excitation 
in time, and encode the distinctive features of time-energy entangled modes.
To see this, we introduce the two-times lesser Green's function~\cite{rammer_2007, loudon_quantum_2000, glauber_quantum_1963} for the photon field
computed at the position $y=0$:
\begin{equation}
    \begin{split}
        G_{\rm pht}(T, & \D t) \equiv \\
        & \braket{\Phi}{\hat{E}^{\dagger}\left(0, T + \frac{\D t}{2} \right)
        \hat{E}\left(0, T - \frac{\D t}{2} \right)}{\Phi}\, .
    \end{split}
    \label{eq:lesser_efield}
\end{equation}
Here,  $\hat{E}(y,t)$ and $\hat{E}^{\dagger}(y,t)$ represent operators in the Heisenberg picture of time evolution, i.e.~ 
$\hat{E}(y,t) \equiv e^{\im \hat{\cal H} t} \hat{E}(y) e^{-\im \hat{\cal H} t}$. The operator $\hat{E}^{\dagger}(y)$ [$\hat{E}(y)$] represents the creation [annihilation] operator of the electric field $\hat{\cal E}(y)$ operator, $\hat{\cal E}(y) \equiv \hat{E}(y)+\hat{E}^{\dagger}(y)$, with
\begin{equation}
    \hat{E}(y)
    =
    \frac{1}{\sqrt{L_y}}
    \sum_{k}\,
    \sqrt{\o_k}\,
    \mathrm{e}^{\mathrm{i}ky}\,
    \aa_{k}
    \label{eq:annihilation_efiled}
    \, .
\end{equation} 

Eq.~(\ref{eq:lesser_efield}) measures the probability amplitude 
for destroying a photon excitation at time $T-\D t/2$ and 
creating it back at time $T+\D t/2$.
By normalizing the Green's function to the joint probability amplitude
for counting electric field excitations at times $T+\D t/2$ and $T-\D t/2$,
we obtain the first-order correlation function~\cite{loudon_quantum_2000, glauber_quantum_1963}
\begin{equation}
    g_{\rm pht}^{(1)} (T,\D t) = 
    \frac{G_{\rm pht}(T,\D t)}
    {\left[ \quave{\hat{E}^{\dagger} \hat{E}}\left(T+ \frac{\D t}{2} \right)
    \quave{\hat{E}^{\dagger} \hat{E}}\left(T- \frac{\D t}{2}\right) \right]^{1/2}}.
    \label{eq:photons_g1}
\end{equation}
When $\left|\gonepht(T,\Delta t) \right| = 1$,  excitations at 
different times  are uncorrelated, namely the probability of 
creating the excitation at time $T+\D t/2$ after having 
destroyed it at time $T-\D t/2$ is exactly equal 
to the joint probability of counting the excitations 	
at initial and final times.

\begin{figure}[t]
    \includegraphics[width=\columnwidth]{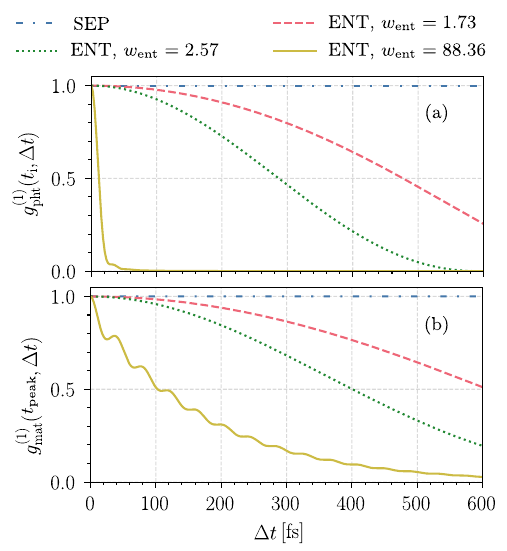}
    \caption{First-order correlation function, see Eq.~(\ref{eq:photons_g1}) and Eq.~(\ref{eq:g1mat}), as a function the 
    propagation time for the photon (a) and matter (b) degrees of freedom. 
    The correlation function is evaluated at the position $y=0$ of the 2D material and  
    for a reference time $T = t_\mathrm{i}$ in panel (a) and $T= t_\mathrm{peak}$ in panel (b). The blue dashed-dotted  line refers to the case of a separable incoming packet with $\tsep\approx232.71~{\rm fs}$. The red dashed line to an entangled packet with 
    entanglement parameter $\went\approx1.73$ and $\tent\approx101.88~{\rm fs}$.
    The green dotted  line to an entangled packet with $\went\approx2.57$ and $\tent\approx79.17~{\rm fs}$. Finally, the yellow solid line to the case of an entangled packet 
    with $\went\approx88.36$ and $\tent\approx2.63~{\rm fs}$. All the incident wave packets, entangled and separable, have a fixed FWHM 
    of approximately $556~{\rm fs}$. }
    \label{fig:fig11}
\end{figure}

In order to highlight the temporal correlation of time-energy 
entangled photons, we compute $g_{\rm pht}^{(1)} (T,\D t)$ in the 
absence of the light-matter interaction.
Fig.~\ref{fig:fig11}(a) reports the quantity 
$\left|\gonepht(T=\ti,\D t)\right|$
for the non-interacting case as a function of the propagation time $\D t$ 
and computed at a reference time $T=\ti$ corresponding to the arrival time of 
the peak of the freely propagating packet at the position $y=0$ of the 2D material.
As expected, for the separable packet we obtain $\left|\gonepht\right|=1$, 
whereas for entangled packets $\left| \gonepht \right|$ deviates 
from $1$ on a time scale set by the entanglement time $\tent.$
For strongly entangled packets, $\gonepht$ rapidly falls 
to zero signaling that, due to time-energy entanglement,
the photon excitations get strongly correlated in time.
 
We now introduce the equivalent Green's functions for the 
matter degrees of freedom and measure the temporal 
correlation induced by the interaction with the entangled 
photons.
The lesser Green's function for the matter degrees of freedom reads as following:
\begin{equation}
    G_{\rm mat} (T,\D t) \equiv \braket{\Phi}{{\bc}\!\left(T+\frac{\D t}{2}\right) 
    {\ba}\!\!\!\left(T-\frac{\D t}{2} \right)}{\Phi},
    \label{eq:Gmat}
\end{equation}
where, as before, the time dependence of the operators 
means Heisenberg evolution.
Since the ground state of coupled light-matter system does not 
coincide with the vacuum---see Appendix~\ref{sec: time evolution}---
the Green's function~(\ref{eq:Gmat}) contains also information about  equilibrium 
temporal correlations, i.e.~in the absence of the driving photon packet.
Of course, due to time translation invariance, 
the equilibrium Green's function depends only on the propagation time $G_{\rm mat}^{\rm equ}(\D t)$.
In order to isolate the modification due the entangled packet, 
we subtract the contribution due to the equilibrium correlations, i.e.~we introduce the quantity 
$\d G_{\rm mat} (T,\D t) \equiv G_{\rm mat}(T,\D t) - G_{\rm mat}^{\rm equ}(\D t),$
and define the first-order correlation function for the matter degrees of freedom as following:
\begin{equation}
    \gonemat(T,\D t)
    \equiv
    \frac{\d G_{\rm mat}(T,\D t)}{\left[\d \quave{\bc \ba}\left(T+\frac{\D t}{2}\right) 
    \d \quave{\bc \ba}\left(T-\frac{\D t}{2}\right) \right]^{1/2}}\, ,
    \label{eq:g1mat}
\end{equation}
where $\d \quave{\bc \ba}(t) \equiv \quave{\bc \ba}(t) 
- \braket{\rm GS}{\bc \ba}{\rm GS}$ represents the number 
of matter excitations on top of the ground state.
In Fig.~\ref{fig:fig11}(b), we show the first-order correlation 
function $\gonemat(T=t_{\rm peak},\D t)$ as a function of the propagation time $\D t$ 
and computed at a reference time $T = t_{\rm peak}$, corresponding 
to the time at which a peak occurs in the  energy absorbed by 
the matter degrees of freedom.
Interestingly, we observe that the first-order correlation 
functions of the matter degrees of freedom acquires the same qualitative 
form of the photonic ones.
For the case of matter degrees of freedom driven by separable packets, the modulus 
of $\gonemat$ is equal to unity,  indicating that the matter excitation 
created by photons does not display temporal correlations.
On the contrary, for entangled incident packets, 
$\left|\gonemat\right|$ deviates from one and decays to zero
on a time scale which decreases upon increasing the entanglement 
parameter $w_{\rm ent}$. This observation provides a clear evidence of the fact 
that, due to the hybridization between matter and 
entangled photons, the matter excitation inherits 
the temporal correlations of the driving pulse.

It is important to stress that the correlation time 
of the matter excitation is not in a quantitative agreement 
with the entanglement time of the incident photon packet.
This can be understood by considering the fact  
that the temporal correlation of the light-stimulated 
matter is the result of the hybridization between the 
matter and photon degrees of freedom.
In particular, the temporal correlations of the incident packets seen in Fig.~\ref{fig:fig11}(a), 
are the result of the superposition of entangled photon modes at all frequencies.
On the contrary, the dynamics in the matter collective 
excitation is dominated by the single resonance frequency $\o_0$
which hybridizes with all the photon modes. 
It is therefore not surprising that the light-induced temporal 
correlations are not in a quantitative agreement with the 
one contained in the driving field.  
We leave for future work a more quantitative understanding 
of the temporal correlations induced by the hybridization between 
matter and entangled photons. 

\section{Conclusions}
\label{sec:conclusions}
We presented a comprehensive study on the effects 
of light entanglement in the light stimulation of a 
collective matter excitation. 
We studied an exactly solvable model in which a single 
partner of a time-energy entangled pair interacts with 
a localized matter mode representing an optically active collective 
excitation of a 2D material.
We showed how the time-energy entanglement in the incident 
photon packets induces qualitative changes in the dynamics of both photon and matter degrees of freedom.

From a purely spectroscopic point of view, our observations  
show that entangled photons act as an extremely 
gentle probe of matter which minimizes the absorbed energy
and maximizes the number of transmitted photons.   
On the other hand, from the point of view of the 
light-induced dynamics, our results show that entanglement driving 
can effectively control the relevant time scales of the light-matter 
interaction process.
For the photons, this is highlighted by the suppression 
of the delay between transmission and absorption peaks. 
For the matter degrees of freedom, entanglement in the driving field suppresses 
the delay between the arrival of the perturbation and the creation 
of the excitation, which is formed on a time scale  
much shorter than the pulse width.
We provide an understanding of these behaviors by 
introducing a representation of entangled photon packets 
in terms of a superposition of trains of short separable pulses. Overall, our results show how driving matter degrees of freedom with temporally-wide entangled 
photons embodies features of ultrashort light 
stimulation.

Our results highlight several future research directions, including,
for example, the study of the role of entangled driving in the creation of 
long-lived light-induced states of matter and the exploration of ultrafast or ballistic 
energy transport by means of absorption of entangled photons~\cite{giacomo_temperonics2021,siemens_ballistic_thermal_transport2010,Campaioli_RMP_2024}.
Moreover, light-induced temporal correlations highlight 
the intriguing perspective of exploiting entangled photons to 
control entanglement in matter degrees of freedom, or, 
 possibly, to create states of matter sharing the entanglement 
properties of the incoming light. 
Finally, the investigation of the correlations between the two entangled photons represents a natural 
framework where the theoretical predictions of this Article can be explored experimentally.

\begin{acknowledgments}
M.P. was supported by the MUR - Italian Ministry of University and Research under the ``Research projects of relevant national interest  - PRIN 2020''  - Project No.~2020JLZ52N (``Light-matter interactions and the collective behavior of quantum 2D materials, q-LIMA'') and by the European Union under grant agreement No. 101131579 - Exqiral and No.~873028 - Hydrotronics. Views and opinions expressed are however those of the author(s) only and do not necessarily reflect those of the European Union or the European Commission. Neither the European Union nor the granting authority can be held responsible for them.
M.D. and D.N.B. were supported by the National Science Foundation under grant No.~DMR-2210186. 
G.M.~acknowledges support by the MUR - Italian Ministry of University and Research through a ``Rita-Levi Montalcini'' fellowship. 
\end{acknowledgments}
    
\FloatBarrier

\appendix

\section{\label{sec: time evolution} Hamiltonian Diagonalization.}
The Hamiltonian $\hat{\mathcal{H}}$, defined in Eq.~(\ref{eq:total_hamiltonian}) of the main text, is quadratic and can therefore be diagonalized by a canonical Bogoliubov transformation.

\subsection{Diagonalization}
Here, we briefly summarize the diagonalization procedure following the algorithm proposed by Colpa in Ref.~\cite{colpa_diagonalization_1978}.

Consider a quadratic Hamiltonian of the form
\begin{equation}
    \hat{\mathcal{H}}
    =
    \frac{1}{2} \, \hat{\bm{\Psi}}^\dagger
    H\,
    \hat{\bm{\Psi}}\, ,
\end{equation}
where
\begin{equation}
    \hat{\bm{\Psi}}
    \equiv
    (\underbrace{\hat{b}, \hat{a}_k, \dots}_\mathcal{N}, \hat{b}^\dagger, \hat{a}^\dagger_k, \dots)^T\, ,
    \quad H^\dagger=H\, .
\end{equation}
We now introduce the operators $\hat{\g}_i$ that diagonalize the Hamiltonian. These operators are defined by the relation
\begin{equation}
    (\hat{\g}_1,\dots,\hat{\g}_{\NN},\hat{\g}^\dagger_1,\dots,\hat{\g}^\dagger_\NN)^T
    \equiv
    \hat{\bm{\O}}
    =
    \mathcal{W} \, \hat{\bm{\Psi}}\, .
\end{equation}
The transformation matrix $\mathcal{W}$ is {\it para-unitary}, satisfying
\begin{equation}
    \mathcal{W}^\dagger \tau^z \mathcal{W} = \tau^z,\quad
    \tau^z \equiv
    \begin{pmatrix}
        \mathbb{1} & 0 \\
        0 & -\mathbb{1} \\
    \end{pmatrix}\, ,
    \label{eq:paraunitary__app}
\end{equation}
which ensures the conservation of the bosonic canonical commutation relations for the $\hat{\gamma}$ operators
\begin{equation}
\big[\hat{\gamma}_a,\hat{\gamma}^\dagger_{a'}\big] = \delta_{a, a'}\, .
\end{equation}
It follows that
\begin{equation}
    \hat{\mathcal{H}}
    =
    \frac{1}{2} \, \hat{\bm{\O}}^\dagger
    \epsilon\,
    \hat{\bm{\O}}\, ,
\end{equation}
where
\begin{subequations}
    \begin{equation}
        (\mathcal{W}^\dagger)^{-1} H \, \mathcal{W}^{-1}
        \equiv
        \epsilon
        \, , \qquad \epsilon_{ab} = \epsilon_a \delta_{a, b}\, ,
    \end{equation}
    \begin{equation}
        \bm{\epsilon}
        =
        (\epsilon_1, \dots, \epsilon_{\mathcal{N}}, \epsilon_1, \dots, \epsilon_{\mathcal{N}})^T\, .
    \end{equation}
\end{subequations}
and $\delta_{a, b}$ is the Kronecker delta.

To find $\mathcal{W}^{-1}$, we first find the Cholesky decomposition
\begin{equation}
    H=K^\dagger K\, .
\end{equation}
Then, we define
\begin{equation}
    \mathcal{U}
    \equiv
    K\mathcal{W}^{-1}\epsilon^{-1/2}
    \, , \qquad
    \mathcal{U}\mathcal{U}^\dagger
    =
    1\, .
    \label{eq:definition_of_Winv__app}
\end{equation}
Furthermore, the unitary matrix $\mathcal{U}$ diagonalizes $K \tau^z K^\dagger$. Indeed,
\begin{multline}
    \mathcal{U}^\dagger K \tau^z K^\dagger \mathcal{U}
    =
    \epsilon^{1/2}\tau^z\epsilon^{1/2}
    \\=
    {\rm diag}(\epsilon_1, \epsilon_2,\dots,\epsilon_{\cal N},-\epsilon_1, -\epsilon_2, \dots, -\epsilon_{\cal N})\, .
\end{multline}
We can now evaluate $\mathcal{W}^{-1}$ from Eq.~(\ref{eq:definition_of_Winv__app}). Furthermore, we define $(\bm{u}_i, \bm{v}_i)^T$ as the $i-{\rm th}$ column of $\mathcal{W}^{-1}$, where $\bm{u}_i$ and $\bm{v}_i$ are ${\cal N}-$component vectors.
Finally, we take the first ${\cal N}$ columns $(\bm{u}_i, \bm{v}_i)^T$ of the matrix $\mathcal{W}^{-1}$ and redefine
\begin{equation}
    \mathcal{W}^{-1}
    \equiv
    \begin{pmatrix}
    \bm{u}_1 & \bm{u}_2 & \dots & \bm{u}_{\cal N} & \bm{v}^\ast_1 & \bm{v}^\ast_2 & \dots & \bm{v}^\ast_{\cal N} \\
    \bm{v}_1 & \bm{v}_2 & \dots & \bm{v}_{\cal N} & \bm{u}^\ast_1 & \bm{u}^\ast_2 & \dots & \bm{u}^\ast_{\cal N} \\
    \end{pmatrix}\, ,
\end{equation}
where $\bm{u}^\ast$ and $\bm{v}^\ast$ are the complex conjugates of $\bm{u}$ and $\bm{v}$.
The matrix $\mathcal{W}^{-1}$ is now para-unitary and diagonalizes $H$.

\subsection{Time evolution}
Introducing the vector
\begin{equation}
    \hbar \bm{f}
    \equiv
    (\epsilon_{1}, \dots, \epsilon_{\mathcal{N}},
    -\epsilon_{1}, \dots, -\epsilon_{\mathcal{N}}
    )^T\, ,
\end{equation}
we write
\begin{equation}
        \big[
    \hat{\mathcal{H}}, \hat{\O}_{i}
    \big]
    =
    -f_i \hat{\O}_{i}\, \Longrightarrow
    \hat{\O}_i(t)
    =
    \mathrm{e}^{-\text{i}f_i t}\,
    \hat{\O}_i(t=0)\, .
\end{equation}

We now show how to use Colpa's algorithm to evaluate the relevant
expectation values.
We are interested in expectation values that contain operators such as
\begin{equation}
    \hat{a}_\mu(t) \, \mathrm{e}^{\text{i}k_\mu y},\,
    \qquad
    \hat{b}(t)\, ,
\end{equation} 
and the corresponding Hermitian counterparts. We introduce 
the $2\mathcal{N}\times \mathcal{N}$ matrix $\mathcal{P}$ defined as
\begin{equation}
    \mathcal{P}_{i \alpha}(t,y)
    \equiv
    \,
    \mathrm{e}^{-\mathrm{i}f_i t}
    \,
    [\mathcal{W}^{-1}]^T_{i \alpha}
    \,
    \mathrm{e}^{\mathrm{i}k_\alpha y}\, .
    \label{eq:P_matrix}
\end{equation}
Using Eq.~\ref{eq:P_matrix}, we can write
\begin{multline}
    \langle \hat{E}^{(-)}(t',y') \hat{E}^{(+)}(t,y) \rangle
    \\=
    \frac{1}{L_y}
    \sum_{\{\alpha,\beta\}=2}^{\mathcal{N}}
    \mathrm{e}^{\text{i}(k_\beta y - k_\alpha y')}
    \sqrt{\omega_\alpha \omega_\beta}
    \langle \hat{a}^\dagger_\alpha(t')\hat{a}_\beta(t) \rangle
    \\=
    \sum_{\{\alpha,\beta\}=2}^{\mathcal{N}}
    \sum_{\{i,j\}=1}^{2\mathcal{N}}
    \frac{\sqrt{\omega_\alpha \omega_\beta}}{L_y}
    \mathcal{P}^\dagger_{\alpha i}(t', y')
    \langle \hat{\O}_i^\dagger \hat{\O}_j \rangle
    \mathcal{P}_{j \beta}(t, y)\, ,
\end{multline}
and
\begin{equation}
    \langle \hat{b}^\dagger(t') \hat{b}(t) \rangle
    =
    \sum_{\{i,j\}=1}^{2\mathcal{N}}
    \mathcal{P}^\dagger_{1, i}(t', 0)
    \langle \hat{\O}_i^\dagger \hat{\O}_j \rangle
    \mathcal{P}_{j, 1}(t, 0)\, ,
\end{equation}
where $\langle \hat{\O}_i \hat{\O}_j \rangle$ is evaluated at the initial time. 
All other relevant expectation values can be calculated in a similar manner.

Details regarding the calculation of the expectation value $\langle \hat{\O}_i \hat{\O}_j \rangle$ at initial time are provided in Appendix~\ref{sec:initialization}.

\section{\label{sec:initialization}Entangled and separable photon state initialization on the ground state}
The interaction Hamiltonian, defined in Eqs.~(\ref{eq:paramagnetic_hamiltonian})-(\ref{eq:diamagnetic_hamiltonian}) of the main text, modifies the ground state of the total Hamiltonian~(\ref{eq:total_hamiltonian}). Consequently the vacuum $|0\rangle$ for the bare annihilation photon and matter operators i.e., $\hat{a}_k|0\rangle=\hat{b}|0\rangle=0$, is no longer an eigenstate of the total Hamiltonian~(\ref{eq:total_hamiltonian}). The first consequence is that the resonant frequency of the material is shifted from $\o_0$ to $\o_\ast$. This effect is discussed in detail in Appendix~\ref{sec:spontaneous_decay_rate}. The second consequence, discussed further in this Appendix, is crucial for accurate numerical simulations.

Initializing the photon state $|\Phi\rangle$ relative to the interacting ground state $|{\rm GS}\rangle$ is necessary to ensure accurate numerical simulations and prevent artifacts in the photon dynamics. Specifically, if $|\Phi\rangle$ is constructed relative to the bare vacuum $|0\rangle$, photon wave packets traveling in opposite directions will be generated at the initial time from the material. This issue arises not from the unitary dynamics itself (which, for a given initial state, is always correct up to numerical precision) but from the fact that $|0\rangle$ is not an eigenstate of the total Hamiltonian. Therefore, even though the unitary dynamics is correct, constructing $|\Phi\rangle$ relative to $|0\rangle$ implies the system is out of equilibrium at the initial time. This is what we term {\it artifacts in photon dynamics}.
Physically, if the initial wave packet is initialized very far from the material, the choice of initialization relative to the bare vacuum or relative to the interacting ground state should be equivalent. However, such an ideal condition can only be achieved approximately in numerical simulations, and we find that only initialization relative to the interacting ground state $|{\rm GS}\rangle$ avoids these simulation artifacts. Here, we explicitly describe how the photon state $|\Phi\rangle$ is initialized relative to $|{\rm GS}\rangle$.

A generic two-photon state $|\Phi\rangle$, defined relative to the interacting ground
state $|\mathrm{GS}\rangle$ of the total Hamiltonian, can be written as
\begin{equation}
    |\Phi\rangle
    \equiv
    \sum_{\mu,\nu}
    \phi_{\mu \nu}\,
    \hat{a}^\dagger_\mu
    \hat{a}^\dagger_{\nu}
    |\mathrm{GS}\rangle
    =
    \sum_{i,j}
    \KK^\dagger_{ij}\,
    \hat{\O}^\dagger_i
    \hat{\O}^\dagger_j
    |\mathrm{GS}\rangle,
\end{equation}
where the matrix $\KK$ is defined as
\begin{equation}
    \KK
    \equiv
    [\mathcal{W}^{-1}]^T\,
    \phi^\dagger\,
    \mathcal{W}^{-1}
    \equiv
    \begin{pmatrix}
        \KK_1 & \KK_2 \\
        \KK_3 & \KK_4 \\
    \end{pmatrix}
    \, ,
\end{equation}
and $\hat{\O}$ is defined as
\begin{equation}
    \hat{\bm{\O}}
    =
    (\hat{\g}_1,\dots,\hat{\g}_{\NN},\hat{\g}^\dagger_1,\dots,\hat{\g}^\dagger_\NN)^T\, ,
\end{equation}
where $|\mathrm{GS}\rangle$ is the vacuum for the $\hat{\gamma}$ operators that
diagonalize the total Hamiltonian and $\mathcal{W}^{-1}$ is the transformation matrix, see App.~\ref{sec: time evolution}.

The state $|\Phi\rangle$ is normalized by imposing $\langle\Phi|\Phi\rangle=1$, which leads to the condition
\begin{equation}
    \langle\Phi|\Phi\rangle
    =
    |\mathrm{Tr}[\KK_2]|^2
    +
    \, \mathrm{Tr}[\KK_1\KK_1^*]
    + \mathrm{Tr}[\KK_1\KK_1^\dagger]
    =
    1\, ,
    \label{eq: state normalization}
\end{equation}
where we have used Wick's theorem.

We now calculate the expectation values
\begin{equation}
    \MM_{\ell n} \equiv \langle\Phi| \hat{\O}^\dagger_\ell \hat{\O}_n |\Phi\rangle
    - \langle\text{GS}| \hat{\O}^\dagger_\ell \hat{\O}_n |\text{GS}\rangle\, .
    \label{eq: Xi expectation value}
\end{equation}
In the following, we assume the Einstein summation convention over repeated indices. We also adopt the notation used in the main text
\begin{equation}
    \langle\quad \rangle_0 \equiv \langle\mathrm{GS}|\quad |\mathrm{GS}\rangle\, ,
    \quad
    \langle \quad \rangle \equiv \langle \Phi| \quad |\Phi \rangle
    -\langle\mathrm{GS}|\quad |\mathrm{GS}\rangle\, .
\end{equation}

Applying Wick's theorem to the expectation value~(\ref{eq: Xi expectation value}) yields
\begin{multline}
    \MM_{\ell n} = \langle \hat{\O}^\dagger_\ell \hat{\O}_n \rangle
    =
    \KK_{ij}\KK_{sp}^\dagger\,
    \langle \hat{\O}_i \hat{\O}_j \hat{\O}^\dagger_\ell \hat{\O}_n \hat{\O}^\dagger_s \hat{\O}^\dagger_p \rangle_0
    \\
    =
    \big(\KK_{ij}\KK_{sp}^\dagger \big)
    \Big(
    \langle \hat{\O}_i \hat{\O}_\ell^\dagger \rangle_0
    \langle \hat{\O}_j \hat{\O}_n \hat{\O}^\dagger_s \hat{\O}^\dagger_p \rangle_{0}
    \\+
    \langle \hat{\O}_j \hat{\O}_\ell^\dagger \rangle_0
    \langle \hat{\O}_i \hat{\O}_n \hat{\O}^\dagger_s \hat{\O}^\dagger_p \rangle_{0}
    +
    \langle \hat{\O}^\dagger_\ell \hat{\O}_s^\dagger \rangle_0
    \langle \hat{\O}_i \hat{\O}_j \hat{\O}_n \hat{\O}^\dagger_p \rangle_{0}
    \\+
    \langle \hat{\O}^\dagger_\ell \hat{\O}_p^\dagger \rangle_0
    \langle \hat{\O}_i \hat{\O}_j \hat{\O}_n \hat{\O}^\dagger_s \rangle_{0}
    \Big)\, ,
\end{multline}
which can be written in the block matrix form
\begin{equation}
    \MM
    =
    \begin{pmatrix}
        \Delta & (\KK_1+\KK_1^T)\, \mathrm{Tr}[\KK_2]^* \\
        (\KK_1^\dagger+\KK_1^*)\, \mathrm{Tr}[\KK_2] & \Delta^*
    \end{pmatrix}\, ,
\end{equation}
where we have introduced the matrix
\begin{equation}
    \Delta \equiv \KK_1\KK_1^\dagger,
    + \KK_1\KK_1^*,
    + \KK_1^T\KK_1^\dagger,
    + \KK_1^T\KK_1^*\, .
\end{equation}
The calculation for an initial separable two-photon Fock state follows directly by setting $\phi_{\mu \nu} = \psi_\mu \varphi_\nu$ in the above expressions.

\section{Spontaneous Decay Rate and Dressed Matter Resonant Frequency\label{sec:spontaneous_decay_rate}}
In this Appendix, we present the analytical and numerical evaluation of the spontaneous decay rate $\G$ defined in Sec.~\ref{sec:light_dynamics} of the main text. We also evaluate numerically the dressed resonance frequency $\o_\ast$ from the transmission coefficient $T(\o)$ defined in Eq.~(\ref{eq:coefficient_transmission}).

\subsection{Analytical calculation}

To calculate the decay rate, we use the Fermi golden rule
\begin{equation}
    \G
    =
    \frac{2\pi}{\hbar}
    |
    \langle\, \text{i} \,|\,
    \hat{\mathcal{\HH}}_{\text{int}}\,
    |\, \text{f}\, \rangle
    |^2
    \rho(E=E_0)\, ,
    \label{eq:fermi_golden_rule__app}
\end{equation}
where $\hat{\mathcal{\HH}}_{\text{int}}$ is a suitable interaction Hamiltonian, $\rho(E=E_0)$ is the density of final states at the transition energy $E_0$, and $|\, \text{i} \,\rangle$ and $|\, \text{f} \,\rangle$ are the initial and final states, respectively.

As far as the interaction Hamiltonian is concerned, we take
\begin{equation}
\hat{\mathcal{H}}_\text{int}
    \equiv
    -\mathrm{i}\hbar \omega_0\tilde{g}
    \sum_{k}
    \frac{1}{\sqrt{\tilde{\omega}_{k}}}
    \big(
    \hat{b}\hat{a}^\dagger_{k}
    +
    \hat{b}^\dagger \hat{a}_{k}
    \big)\, ,
    \label{eq_app:interaction_Hamiltonian}
\end{equation}
which connects initial and final states with the same energy. The interaction Hamiltonian (\ref{eq_app:interaction_Hamiltonian}) can be obtained from the full light-matter interaction $\hat{\cal H}_{{\rm mat}-{\rm pht}}$ in the rotating-wave approximation~\cite{loudon_quantum_2000}.

To calculate the spontaneous decay rate, we consider the transition from the initial state
\begin{equation}
    |\,\mathrm{i}\, \rangle
    \equiv
    |\,0_{\text{photons}};\,1_{\text{phonons}}\,\rangle\, ,
\end{equation}
which contains a single phonon with energy $E_0$, to the final state
\begin{equation}
    |\,\mathrm{f}\, \rangle
    \equiv
    |\,1_{\text{photons, $\o_0$}};\,0_{\text{phonons}}\,\rangle\, ,
\end{equation}
which contains a single photon with energy $\hbar \o_0=E_0$.
It follows that
\begin{equation}
    |\langle \, \mathrm{i}\,|\,
    \hat{\mathcal{H}}_\text{interaction}\,
    |\, \mathrm{f}\, \rangle
    |^2
    =
    (\hbar \omega_0 \tilde{g})^2
    =
    \frac{\hbar^2 \omega_0 \tilde{q}^2 \alpha \pi c}{L_y}\, .
    \label{eq:S_matrix_elements}
\end{equation}

We now calculate the density of states $\rho(E)$ for the photons.
If $\mathfrak{N}_{\text{pht}}$ is the number of photon modes, we can write
\begin{equation}
    \mathrm{d}\mathfrak{N}_{\text{pht}}
    =
    \frac{\mathrm{d}k_y}{2\pi/L_y}
    =
    \frac{L_y}{\hbar c \pi} \mathrm{d}E
    \equiv
    \rho(E)\mathrm{d}E\, ,
\end{equation}
which implies
\begin{equation}
    \rho(E)
    =
    \frac{L_y}{\hbar c \pi}\, .
    \label{eq:density_of_states}
\end{equation}
Finally, substituting Eq.~(\ref{eq:S_matrix_elements}) and Eq.~(\ref{eq:density_of_states}) into the Fermi golden rule Eq.~(\ref{eq:fermi_golden_rule__app}), gives the spontaneous decay rate
\begin{equation}
    \Gamma
    =
    2\omega_0 \pi \tilde{q}^2 \alpha
    \approx
    6.97~{\rm THz}\, ,
\end{equation}
where in the last equation we used $\hbar \o_0=100~{\rm meV}$ and $\tilde{q}=1$.

\subsection{Numerical estimation}
\begin{figure}
    \includegraphics[width=\columnwidth]{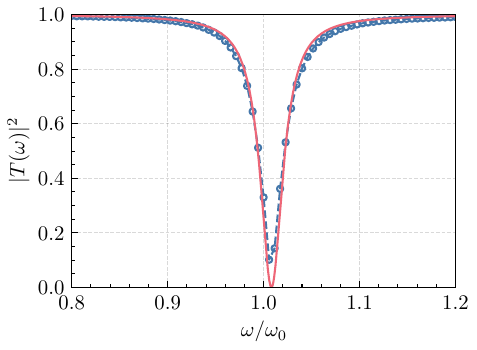}
    \caption{Absolute square value of the coefficient transmission. 
    The blue points are the numerical data, while the solid orange line is evaluated from a 
    fit with the model function Eq.~(\ref{eq:app_Tk})}
    \label{fig:app_fig1}
\end{figure}
Here, we estimate the spontaneous decay rate from the transmission coefficient $T(\o)$ defined in Eq.~(\ref{eq:coefficient_transmission}).
In Fig.~\ref{fig:app_fig1}, we show numerical data for the squared magnitude of the transmission coefficient. The data are fitted using the model function
\begin{equation}
    T(\omega)
    =
    \frac{\o-\o_*}
    {\o-\o_*+\im \G/2}\, ,
    \label{eq:app_Tk}
\end{equation}
whose form is based on Eq.~(\ref{eq:coefficient_transmission}) in the main text.
The fit yields the parameters $\o_* = (1.0082 \pm 0.0001)\, \o_0$ and $\G = (4.31 \pm 0.03)\, \mathrm{THz}$, evaluated assuming $\hbar\o_0 = 100~{\rm meV}$.

\section{\label{sec:entanglement_entropy} Entanglement Entropy}
Here, we briefly explain how to compute the von Neumann entanglement entropy introduced in Sec.~\ref{sec:model}. More details can be found in Ref.~\cite{law_continuous_2000}.

Consider a generic two-mode Fock state
\begin{equation}
    |\Phi\rangle \equiv \sum_{\mu\nu} \phi_{\mu\nu}\, \hat{a}^\dagger_\mu \hat{a}^\dagger_\nu \,|0\rangle\, ,
    \qquad
    \langle\Phi|\Phi\rangle=1\,.
\end{equation}
We define two matrix kernels
\begin{equation}
    K_1 \equiv \bm{\phi} \bm{\phi}^\dagger\, , \qquad
    K_2 \equiv \bm{\phi}^\dagger \bm{\phi}\, ,
\end{equation}
and introduce the orthonormal eigenvectors $\psi_\mu^{(n)}$ and $\varphi_\mu^{(n)}$, satisfying
\begin{equation}
    K_1 \psi_\mu^{(n)} = \lambda_n \psi_\mu^{(n)}\, , \qquad
\, 		K_2 \varphi_\mu^{(n)} = \lambda_n \varphi_\mu^{(n)}\, ,
\end{equation}
where the eigenvalues $\lambda_n$ are real and non-negative. The Schmidt modes are then introduced as
\begin{equation}
    \hat{A}_n \equiv \sum_\mu \psi_\mu^{(n)} \hat{a}_\mu\, , \qquad
    \hat{B}_n \equiv \sum_\mu \varphi_\mu^{(n)} \hat{a}_\mu\, ,
\end{equation}
so that the two-mode Fock state can be written as a sum of factorized states (Schmidt decomposition)
\begin{equation}
    |\Phi\rangle
    =
    \sum_n \lambda_n\, \hat{A}^\dagger_n \hat{B}^\dagger_n\, |0\rangle\, .
\end{equation}
The von Neumann entanglement entropy is then evaluated as
\begin{equation}
    E_{\text{N}} = - \sum_n \lambda^2_n \log(\lambda^2_n)\, .
\end{equation}

\section{Analytical Model\label{sec:analytical_model}}
In this Appendix, we derive the equations for the
analytical model introduced in Sec.~\ref{sec:light_dynamics}.

For definitions of special functions and standard integrals
and sums, we refer to Ref.~\cite{gradstejin_table_2009}.
Furthermore, we use the notation $\partial^j_xf\equiv\frac{\partial^j f}{\partial x^j}$ to denote
the $j$-th derivative of a function $f(x)$ with respect to $x$.
\subsection{Normalization constant}
We first calculate the normalization constant $N$ that appears in the function $\psi(\o)$ defined in Eq.~(\ref{eq:psielementary}) of the main text. To calculate $N$, we focus on the non-interacting case, which implies $T(\o)=1$.
In this limit, the wavefunction $\Psi_0(\t)$---see Eq.~\eqref{eq:spectral_representatio_Psi}---simplifies to
\begin{equation}\label{eq:Psi0noninteracting_in_appendix}
    \Psi^{(0)}_0(\t)=N\,\mathrm{e}^{-\t^2/2\t_+^2}\, .
\end{equation}
where the superscript $(0)$, like in the main text, refers to the limit $T(\o)\to1$.

The $j$-th wave function $\Psi_j(\t)$ is proportional to the $j$-th momentum $\llangle \o^j \rrangle(\t)$---both in the interacting and non-interacting case---which can be written as
\begin{equation}
    \llangle\o^j\rrangle(\t)
    =
    \im^j \partial^j_\t \Psi_0(\t)\, .
\end{equation}
We define the dimensionless variable $\eta \equiv \tau/\tau_+$, which implies
\begin{equation}
    \partial^j_\tau \Psi_0^{(0)}
    =
    \tau_+^{-j}\, \partial^{(j)}_\eta \Psi_0^{(0)}
    =
    N
    \mathrm{e}^{-\eta^2/2}
    \frac{(-1)^j}{\t_+^j\, 2^{j/2}}\mathrm{H}_j\Big(\frac{\eta}{\sqrt{2}}\Big)\, ,
\end{equation}
where we used the definition of the Hermite polynomials
\begin{equation}
    \mathrm{H}_j(\eta)
    =
    (-1)^j
    \mathrm{e}^{\eta^2}
    \partial^j_\eta
    \mathrm{e}^{-\eta^2}\, .
\end{equation}
For the elementary photon densities $n^{(0)}_j(\t)$ defined in Eq.~(\ref{eq:nj_transmitted}) of the main text, it thus follows
\begin{equation}
    n^{(0)}_j(\t)
    =
    |\Psi_j^{(0)}(\t)|^2
    =
    \mathrm{e}^{-\nicefrac{\t^2}{\t_+^2}}
    \mathrm{H}_j^2\Big(\frac{\t}{\sqrt{2}\tau_+}\Big)
    \Big(
    \frac{\t_-}{\t_+}
    \Big)^{2j}
    \frac{N^2}{j!\,2^j}\, ,
\end{equation}
where
\begin{equation}
    \Psi_j^{(0)}(\t)
    =
    \frac{\t_-^j}{\sqrt{j!}}
    \llangle\o^j\rrangle^{(0)}(\t)\,,
\end{equation}
as defined in Eq.~\eqref{eq:pisj_omegaj} of the main text.

We define
\begin{multline}
    I_j
    \equiv
    \int \mathrm{d}\eta
    \,
    (\partial^j_\eta \Psi_0^{(0)})(\partial^j_\eta \Psi_0^{(0)})
    \\=
    (-1)^j
    \int \mathrm{d}\eta
    \,
    (\partial^{2j}_\eta \Psi_0^{(0)}(\eta)) \Psi_0^{(0)}(\eta)
    \\=
    \frac{N^2(-1)^j\sqrt{2}}{2^j}
    \int
    \mathrm{d}\eta\,
    \mathrm{e}^{-2\eta^2}
    \mathrm{H}_{2j}(\eta)\, ,
\end{multline}
where the second equality follows from repeated integration by parts.
Using the relation
\begin{multline}
    \sum_{j=0}^{+\infty}
    \int
    \mathrm{d}\eta\,
    \mathrm{e}^{-2\eta^2}
    \mathrm{H}_j(\eta)
    \frac{t^j}{j!}
    =
    \int
    \mathrm{d}\eta\,
    \mathrm{e}^{2\eta t-t^2-2\eta^2}
    \\=
    \mathrm{e}^{-t^2/2}\sqrt{\frac{\pi}{2}}
    =
    \sqrt{\frac{\pi}{2}}
    \sum_{j=0}^{+\infty}
    \frac{(-1)^j}{2^j j!}
    t^{2j}
\end{multline}
we find
\begin{equation}
	\int
	\mathrm{d}\eta\,
	\mathrm{e}^{-2\eta^2}
	\mathrm{H}_{2j}(\eta)
	=
	\sqrt{\frac{\pi}{2}}
	\frac{(-1)^j (2j)!}{2^j j!}
\end{equation}
and thus
\begin{equation}
    I_j
     =
    \frac{N^2(-1)^j\sqrt{2}}{2^j}
    \int
    \mathrm{d}\eta\,
    \mathrm{e}^{-2\eta^2}
    \mathrm{H}_{2j}(\eta)
    =
    \frac{N^2\sqrt{\pi}(2j)!}{2^{2j} j!}\,.
\end{equation}
We can now calculate the total number of photons carried by the $j$-th elementary photon density---Eq.~(\ref{eq:Njzero})---by performing the integration
\begin{equation}
    N_j^{(0)}
    =
    c\tau_+
    \int
    \mathrm{d}\eta\,
    n_j^{(0)}(\eta)
    =
    N^2c\t_+\sqrt{\pi}
    \binom{2j}{j}
    \Big(\frac{\tau_-}{2\t_+}\Big)^{2j}
    \, .
    \label{eq:number_og_photon_jth__app}
\end{equation}
The total number of photons is obtained by summing Eq.~(\ref{eq:number_og_photon_jth__app}) over $j$ i.e.,
\begin{equation}
	N^{(0)}
	=
	\sum_{j=0}^{+\infty}
	c\tau_+
	\int
	\mathrm{d}\eta\,
	n_j^{(0)}(\eta)
	=
	\frac{N^2 c\t_+ \sqrt{\pi}}{\sqrt{1-(\tau_-/\tau_+)^2}}\, ,
\end{equation}
with $(\tau_-/\tau_+)^2<1$. We normalize the state to contain a single photon ($N^{(0)}=1$) by imposing
\begin{equation}
	N^2
	=
	\sqrt{1-(\tau_-/\tau_+)^2}/(c\t_+ \sqrt{\pi})\, .
\end{equation}

\subsection{Transmitted wave packet}
We now focus on the transmitted separable wave
packet. We evaluate $\Psi_0(\t)$ defined in Eq.~(\ref{eq:psielementary}) of the main text, including the effect of the transmission coefficient, i.e.
\begin{multline}
    \Psi_0(\t)
    =
    \frac{N\t_+}{\sqrt{2\pi}}
    \int_{-\infty}^{+\infty}
    \mathrm{d}\o\,
    \mathrm{e}^{-\frac{\o^2\t_+^2}{2}-\im \o\t}\,
    \frac{\o}{\o+\mathrm{i}\G/2}
    \\=
    N
    \mathrm{e}^{-\frac{\t^2}{2\t_+^2}}
    -
    \im N\G
    \int_{-\infty}^{+\infty}
    \mathrm{d}\o\,
    \mathrm{e}^{-\frac{\o^2\t_+^2}{2}-\im \o\t}\,
    \frac{1}{\o+\mathrm{i}\G/2}\, .
    \label{eq:f0}
\end{multline}
We introduce the integral representation of the denominator
\begin{equation}
    \frac{\mathrm{i}}{\o+\mathrm{i}\G/2}
    =
    \int_{-\infty}^{+\infty}
    \mathrm{d}\l \,
    \mathrm{e}^{-\text{i}\l\o+\G\l/2}\,
    \Theta(-\l)\, ,
    \label{eq:integral_representation}
\end{equation}
where $\Theta(\l)$ is the Heaviside distribution. Substituting Eq.~(\ref{eq:integral_representation}) into Eq.~(\ref{eq:f0}), and performing the Gaussian integral over $\o$, we rewrite
\begin{multline}
    -
    \mathrm{i}\frac{\G}{2}
    \int_{-\infty}^{+\infty}
    \mathrm{d}\o\,
    \mathrm{e}^{-\o^2\t_+^2/2-\text{i}\o\t}
    \frac{1}{\o+\mathrm{i}\G/2}
    \\=
    -
    \frac{\G}{2}
    \int_{-\infty}^{+\infty}
    \mathrm{d}\l \,
    \int_{-\infty}^{+\infty}
    \mathrm{d}\o\,
    \Theta(-\l)\,
    \mathrm{e}^{-\frac{\o^2\t_+^2}{2}-\im \o(\t+\l)+\frac{\G\l}{2}}
    \\=
    -
    \frac{\G}{2}
    \int_{-\infty}^{+\infty}
    \mathrm{d}\l \,
    \mathrm{e}^{\frac{\G\l}{2}-\frac{(\l+\t)^2}{2\t_+^2}}\,
    \Theta(-\l)
    \\=
    \sqrt{\frac{\pi}{2}}
    \frac{\G}{2}\t_+
    \mathrm{e}^{\frac{\G^2\t_+^2}{8}-\frac{\G\t}{2}}
    \left(
    \! \mathrm{Erf}
    \left[
    \frac{\G\t_+}{2\sqrt{2}}-\frac{\t}{\sqrt{2}\t_+}
    \right]-1
    \! \right)
    \, ,
\end{multline}
which implies
\begin{multline}
    \Psi_0(\t)
    =
    N
    \bigg(
    \mathrm{e}^{-\t^2/2\t_+^2}
    \\+
    \sqrt{\frac{\pi}{2}}\,
    \G\t_+\,
    \mathrm{e}^{\G^2\t_+^2/2-\G\t}\,
    \big(\mathrm{Erf}[(\G\t_+-\t/\t_+)/\sqrt{2}]-1)\big)
    \bigg)
\end{multline}
The transmitted elementary photon density $n_j(\t)$ can then be evaluated as
\begin{equation}
    n_j(\t)
    =
    |\Psi_j(\t)|^2
    =
    \Big|
    \llangle \o^j \rrangle(\t)
    \Big|^2
    \frac{\t_-^{2j}}{j!}
    =
    \Big|
    \partial_\t^j
    \Psi_0
    \Big|^2
    \frac{\t_-^{2j}}{j!}\, .
\end{equation}

\end{document}